\documentstyle[11pt,a4]{article}
\begin{document}

\centerline {\LARGE \bf Phase transitions in open quantum systems}

\vspace*{1cm}

\centerline{\bf \Large C.~Jung$^{1,2}$, M.~M\"uller$^{1,3}$ and 
I.~Rotter$^{1,4}$
}
\vspace*{.3cm}

{\it
\begin{center}
$^1$ Centro Internacional de Ciencias, Cuernavaca, Mexico\\
$^2$ Instituto de Matematicas, Unidad Cuernavaca, UNAM\\
Apdo. postal 273-3, 62151 Cuernavaca, Mexico\\

$^3$ Max-Planck-Institut f\"ur Physik komplexer Systeme\\
N\"othnitzer Str. 38, D-01187  Dresden, Germany \\
$^4$
Technische Universit\"at Dresden, Institut f\"ur Theoretische Physik,\\
D-01062 Dresden, Germany, and  \\
 Max-Planck-Institut f\"ur Physik komplexer Systeme\\
N\"othnitzer Str. 38, D-01187  Dresden, Germany \\

\end{center}}

\vspace*{.5cm}

05.70.Fh, 03.80.+r, 64.60.-i, 03.65.-w

\vspace*{1.5cm}

\begin{abstract}

We consider the behaviour of open quantum systems in dependence on 
 the coupling  to 
one decay channel by introducing the coupling parameter $\alpha$
being proportional to the average degree of overlapping.  
Under critical conditions, a
 reorganization of the spectrum takes place which
 creates a bifurcation
of the time scales with respect to the lifetimes of the resonance states.
We derive analytically the conditions under which  the 
reorganization 
process  can be understood as a second-order phase transition
and illustrate our results by numerical investigations.
The conditions are fulfilled e.g. for a picket fence with equal coupling
of the states to the continuum.
Energy dependencies within the system are included.
We consider 
also  the generic case of an
unfolded Gaussian Orthogonal Ensemble.
In all these cases, the reorganization of the spectrum occurs 
at the critical value $\alpha_{\rm crit}$ of the control parameter
globally over the
whole energy range of the spectrum. All states  act 
cooperatively.

\end{abstract}

\vspace{1cm}
\newpage

\section{Introduction}

Recently, the properties of open quantum systems have been
studied
with a renewed interest  in the framework of different approaches.
Mostly discussed is the  restructuring of the systems taking
place at high level density under critical conditions and  the  resulting
formation of different time scales in terms of  lifetimes
of  resonance states. 
The reorganization occurs if
the  degree $\bar \Gamma / \bar D $
 of overlapping reaches a critical value
($\bar \Gamma$ is the average width obtained by averaging over {\it all} $M$
resonance states in a certain energy region
 and $\bar D$ is the mean level distance).
It is investigated for resonance 
phenomena  in
 nuclei \cite{ro91,soze,diharo,ismuro,levrep,sorosamu}, 
atoms \cite{frwi,atom} and  molecules
 \cite{mole}. 
In the meantime it has been considered  also in other systems
such as e.g.  quantum dots \cite{fyso} and microwave billards
\cite{seba}. 
The number $M$ of resonance states is usually
much larger than the number $K$ of open decay channels. 
\\

A powerful method in describing the properties of an open
quantum system is  the projection operator technique.
 It allows us to investigate, in a direct manner,  the corrections 
to the many-particle states  in a subspace 
of the full Hilbert space which arise from  the 
coupling to the orthogonal  subspace.
Originally the idea of using the projection operator technique arose 
in
nuclear physics. The  influence of 
discrete states as well as closed decay channels
($Q $ subspace) onto the wavefunctions of the
open decay channels ($P $ subspace) 
 is considered
in the Feshbach unified theory of nuclear reactions \cite{fesh}. 
This phenomenological model
 describes well the properties of nuclei at an
excitation energy of about 10 MeV
by  employing statistical assumptions on the  states of the $Q$ subspace. 
 In contrast to this, the ground and 
low-lying states of nuclei are described well by the  wavefunctions 
of the discrete states ($Q$ subspace), 
 considering the influence of the decay channels ($P$ subspace) 
in an approximated manner.
\\

 The properties of an {\it open} quantum system are described 
well by using
the latter division of the whole function space: 
the $Q$ subspace contains the discrete states of the system
 while the $P$ subspace
consists of  open as well as closed decay channels \cite{ro91}. 
In studying the restructuring, we are interested in the properties of the
 states of the $Q$
subspace modified by their coupling to the $P$ subspace playing the role of 
an environment. 
This method can be used for a wide class of open quantum systems
\cite{mawe}.
 \\

 The question whether this
restructuring may be considered as a phase transition of second order is
put in \cite{diharo} but not considered in detail up to now. A possible
analogy to the formation of laser light is investigated numerically in
\cite{ismuro}.  As in the case of the laser,
a
control 
parameter $\alpha$ can be defined which is proportional 
to $\bar \Gamma / \bar D$.
  The information entropy  changes rapidly  in a
relatively small region of the control 
parameter in both the laser
\cite{haken} and the open quantum system \cite{ismuro}.  \\

In other investigations \cite{levrep} it 
was realized that the avoided
crossing of two neighbouring  
resonance states  which are coupled to one common
  channel is the basic process of the restructuring observed globally in
  the system. As soon as two resonances start to overlap each
  other, their interaction via the continuum can no longer be neglected. As
  a function of the coupling to a certain decay channel
  the two resonances approach each other in energy up to a certain minimum
   distance  in the complex energy plane at $\alpha = \alpha_{{\rm crit}}$.  
The avoided crossing is
  reflected in the wavefunctions of the two resonance states.
The biorthogonality reaches its maximum if
 $\alpha \to \alpha_{{\rm crit}}$,
   and vanishes
   if $\alpha \to 0$ and $\alpha \to \infty$ \cite{levrep}.    As a
  function of further increasing $\alpha > \alpha_{{\rm crit}}$, the width 
of one of the two resonance states decreases (resonance trapping) 
  while the width of the other one increases further. 
  \\

The local resonance trapping can explain, indeed, the global
restructuring of the quantum system under critical conditions. 
 It determines also, as will be shown in this
paper, whether the restructuring of the system takes place collectively
with the simultaneous participation of {\it all} basis states
or successively 
by individual trapping of resonance states.  \\

In the following, we will investigate this question in detail. In sect.~2,
we write down the basic equations used in the 
paper. The model is formulated and the characteristic polynomial is given.
The Hamiltonian is non-hermitian and its eigenfunctions are, generally,
bi-orthogonal. In section 3, the properties of a system with  picket-fence
distributed levels coupled with the same strength to one decay channel is 
investigated in detail. The study is performed analytically for the limiting
case of an infinite number of states as well as for a finite number. The
results are illustrated by numerical 
calculations. Finally, the results
obtained are discussed and
identified with characteristic features of a second-order
phase transition. The process of formation of a collective
state aligned with the decay channel is discussed in detail. Its 
wavefunction is
coherently mixed in the wavefunctions of all basis states including those
states which are {\it not} overlapped by it.
\\

The results obtained in section 3 are underpinned 
in section 4 by considering 
some 
other level and coupling-strength distributions being more realistic than those
in section 3. The study is performed both analytically and numerically.
General conditions for the appearance of a second-order phase transition 
are formulated analytically and illustrated by the results of numerical
calculations. As a special case, the sharpness of a
phase transition is shown to be distorted 
by an imaginary part in the coupling term. The results are summarized
and discussed in the last section. 
\\

\section{Basic equations}

\subsection{The Hamiltonian of an open quantum system}

Let us consider the Hamilton operator 
\begin{eqnarray}
H = H_0 + \hat V 
\label{eq:hamtot}
\end{eqnarray}
of a many-particle system where 
$H_0$ describes the mean-field, i.e. the motion of  
 the particles in a
finite depth potential 
and $\hat V$ is the operator of the two-particle
residual interaction. 
A convenient method to solve the Schr\"odinger equation $(H - E)\Psi = 0$ 
in the full Hilbert space of discrete and continuous states is to 
use the projection operator technique introduced by Feshbach \cite{fesh}. 
Here the whole function space
is divided into two subspaces by using the projection operators 
\begin{eqnarray}
\hat Q&=&\sum_{k=1}^M |\Phi_k\rangle\langle\Phi_k| 
\nonumber\\ 
\hat P&=&\sum_c\int_{E_c}^{\infty}dE'|\xi_c(E')\rangle\langle\xi_c(E')|
\end{eqnarray} 
with ${\hat P} + {\hat Q} = 1$,  
where  $Q$ contains the many-particle discrete states $|\Phi_k\rangle$ 
being solutions of $({\hat Q}H{\hat Q} - E_k) \; |\Phi_k\rangle = 0$, 
and $P$ the many-particle scattering states $|\xi_c(E)\rangle$
being solutions of $(\hat PH\hat P - E)\; |\xi_c(E)\rangle = 0$. 
The total Hamiltonian acting on the full Hilbertspace is split
into
four terms: $H=\hat QH\hat Q+\hat QH\hat P+\hat PH\hat Q+\hat PH\hat P$.\\

We are interested in the properties 
of the Hamiltonian
of the open quantum system, which acts on the $Q$ subspace 
{\it and} carries the influence of the $P$ space.
The derivation of this Hamilton operator can be found in
\cite{ro91,mawe},
\begin{eqnarray}
{H}^{\rm eff}_{QQ}(E) = \hat QH\hat Q + \hat QH\hat P \cdot G_P^{(+)}(E) \cdot \hat PH\hat Q \; .
\label{eq:ham0}
\end{eqnarray}
It consists of two terms. The first one ($\hat QH\hat Q$)
describes the behaviour of the closed
system of discrete states which includes the configurational mixing due to 
the two body residual interaction,
but does not take into account the coupling to the decay channels.
The second term gives the correction due to the coupling
of the two subspaces and contains the 
propagator in the $P$-subspace $G_P^{(+)}(E)=\hat P[E+i\eta-\hat PH\hat P]^{-1}\hat P$. \\

Due to this propagator the effective Hamiltonian is energy dependent and non-hermitian.
Its complex eigenvalues $\lambda_k(E)={\cal E}_k(E) -
\; i/2\;\; \Gamma_k(E)$ give the poles of the 
resonance part of the scattering matrix 
\begin{eqnarray}
S^{{\rm res}}_{cc'} = i\sum_{k=1}^M \frac{\gamma_{kc}(E)\gamma_{kc'}(E)}{E-\lambda_k(E)}
\label{eq:smat}
\end{eqnarray}
where $\gamma_{kc}(E)=1/\sqrt{2\pi}\; \langle\xi_c(E)|\hat V|\Phi_k\rangle$ is the transition matrix element
between a bound and a scattering state. Thereby, 
the complex
eigenvalues $\lambda_k$ 
get a concrete
physical interpretation as the energy positions ${\cal E}^{{\rm res}}_k={\cal
E}_k({\cal E}^{{\rm res}}_k)$ 
and total decay widths 
$\Gamma^{{\rm res}}_k=\Gamma_k({\cal E}^{{\rm res}}_k)$ of a resonance state
\cite{ro91,mawe}. 
The ${\cal E}_k$ differ 
usually from the corresponding 
eigenvalues $E_k$ of ${\cal H}^0 \equiv \hat QH\hat Q$, i.e. from the  energies of the states 
of the unperturbed system.
So the external coupling to the decay channels causes 
not only the finite lifetime of the states but generally
also an energy shift.
\\

In the following, we will restrict ourselves to an energy region in which the
energy dependence of the Hamiltonian is small in spite of a large number $M$
of states lying in it. Further, we consider a small number $K$ of decay
channels which are all open and not coupled among themselves. 
Then, the effective Hamiltonian  
 ($\ref{eq:ham0}$)
in the $Q$ subspace is, to a good approximation,
\begin{eqnarray}
{\cal H} = {\cal H}^0 -i\alpha VV^+
\label{eq:ham}
\end{eqnarray}
where $VV^+$ is a hermitian operator if we consider time reversal invariance.
As in Eq.~(\ref{eq:ham0}) the first term ${\cal H}^0$ describes the internal
structure of the 
unperturbed
system in the $Q$ subspace.
The second term $ i \alpha V V^+$ follows from $ \hat QH\hat P 
\cdot G_P^{(+)}(E)
 \cdot \hat PH\hat Q$ and
 describes the coupling between the two subspaces. 
The parameter $\alpha$, assumed mostly to be real,  characterizes the
mean coupling strength between discrete and continuous states.       
\\

The Hamiltonian  ($\ref{eq:ham}$) is used successfully for the description 
of resonance states in  nuclei \cite{mawe} and molecules \cite{mole}. 
Nowadays, it is applied 
also to the description of resonance phenomena in other systems 
such as e.g. quantum dots \cite{fyso} and microwave billards 
\cite{seba}. 
\\

The rank of ${\cal H}^0$ is equal to the number $M$ of states considered. 
Its non-diagonal matrix elements describe the configurational mixing of the discrete states. 
The coupling matrix $V$ is  a $K \times M$ matrix 
 if the number of open decay channels is equal to $K$.
The element $V_i^c$ of $V$ describes the coupling of the discrete
state $i$ to the channel $c$;  $\; i = 1,...,M; \; c = 1,...,K$.
Thus, the rank of $V V^+$ is  $K$. 
\\

As long as $\alpha$ is small, the second term of the Hamiltonian ${\cal H}$
can be considered as a small perturbation of ${\cal H}^0$. This condition is
always fulfilled if the  average width  $\bar\Gamma $ is 
much smaller than the average distance $\bar D$
between  neighbouring 
resonance states.
In this case, the non-diagonal matrix elements of ${\cal H}$ are
small, the individual resonances are isolated.
Their positions and widths obtained from the eigenvalues $\lambda _i$ 
of ${\cal H}$ differ only slightly from the real and
imaginary parts, respectively, of the diagonal matrix
elements of ${\cal H}$. \\

In the opposite case of large $\alpha$, the matrix $V V^+$ determines the
behaviour of the system. Then, the rank of ${\cal H}$ is  given by $K$.
That means,  $M - K$ states are almost decoupled 
from the continuum of decay channels and become long lived (trapped) while 
 $K$ states take almost the whole 
coupling strength: 
$\sum_{i=1}^K \Gamma_i / 2 \approx \Im \{Tr
({\cal H}) \}$ and
 $\sum_{i=K+1}^N \Gamma_i \approx 0 $.
Therefore, two different time scales arise at large $\alpha$, see e.g.
\cite{ro91,soze,ismuro,levrep,mole}.\\

Thus, a reorganization  in the open quantum system takes place in the
transition from small coupling parameters  $\alpha$ to large ones 
when $M \gg K$. 
In the following, we will investigate the question whether and under which 
conditions the reorganisation
of the open quantum system 
can be understood as a phase transition
 in the limit $M \to \infty$.
We restrict ourselves to the case with one open decay channel ($K=1$).\\

\subsection{The characteristic polynomial}

 We consider a system with $M=2N+1$ states coupled to one 
common decay
 channel ($K=1$). The unperturbed eigenvalues of ${\cal H}^0$ are denoted by
 $E_k$, $k\in\{-N,...,N\}$, so that $E_j<E_k$ if $j<k$ 
(without degeneration).
The centre of the spectrum is assumed to be at $E_0 = 0$ 
without loss of generality. The coupling vector 
will be denoted by $V=(v_{-N},...,v_{-1},v_{0},v_{1},...,v_{N})$.
\\

Due to $K=1$,   
 all column and row vectors, respectively, of  $VV^+$ are linearly dependent.
Substracting $v_k$ times the row $0$ from the row $k$, one gets the following
 expression for the characteristic polynomial, 
{\small
\begin{displaymath}
P_{N}(\lambda) = 
\left |
\begin{array}{cccccccc}
E_{-N}-\lambda& 0 & 0  & \ldots &  \lambda v_{-N}   & 0 & \ldots &   0     \\

   0   & E_{-N+1}-\lambda & 0 & \ldots &  \lambda v_{-N+1} & 0 & \ldots &   0     \\

\vdots &     & \ddots &         &  \vdots  & \vdots &        &  \vdots \\

       &     &      &         &         &        &        &   \\
   0   &  \ldots & 0 & E_{-1}-\lambda & \lambda v_{-1} & 0 & \ldots  & 0    \\

-i\alpha v_{-N} & -i\alpha v_{-N+1} &   \ldots &   -i\alpha v_{-1} & -i\alpha v_0 - \lambda & -i\alpha v_1 & \ldots &   -i \alpha v_N \\

  0   &  & \ldots  &   & \lambda v_{1} & E_1 -\lambda & \ldots &   0   \\
\vdots &     & \ldots &   &  \vdots  &   & \ddots &  \vdots \\
  0   &   & \ldots  &   &  \lambda v_{N} &  0  & \ldots &   E_N -\lambda \\
\end{array}
\right | = 0
\end{displaymath}
}
which can be written as
\begin{equation}
P_N(\lambda) = \prod_{k=-N}^{N} (E_k - \lambda) - i\alpha \cdot \sum_{k=-N}^{N}
|v_k|^2 \cdot \prod_{j=-N,j\ne k}^N (E_j - \lambda) = 0 \; .
\label{eq:CP}
\end{equation}
Eq.~($\ref{eq:CP}$) can be proven by induction.\\

According to Eq.~($\ref{eq:CP}$),  $P_N (\lambda)$ is the sum of two 
polynomials, 
\begin{equation}
P_N(\lambda) = Q_N(\lambda) - i\alpha \cdot R_N(\lambda),
\label{eq:CPstrukt}
\end{equation}
where $Q_N$ is of the order $2N+1$ and $R_N$ of the order $2N$. 
If  $|v_k|^2 = 1 \;\;\; \forall k$,
the $Q_N$ and $R_N$ are related in a simple manner,
\begin{eqnarray}
R_N = - \frac{d}{d\lambda} Q_N \; .
\label{eq:PFzus}
\end{eqnarray}
In the limit  $\alpha=0$, we find
$\lambda_k={\cal  E}_k=E_k \;\; \forall k$,  i.e. the eigenvalues of 
${\cal H}$ are equal to those of ${\cal H}^0$ (according to the definition 
of the parameter $\alpha$).\\

The  limit of large coupling strength $(\alpha \to \infty)$ can be obtained when
  we rewrite the characteristic polynomial 
(\ref{eq:CP}) as
\begin{eqnarray}
P_N(\lambda) = i\alpha \prod_{k=-N}^{N} (E_k - \lambda) \left [
\frac{1}{i\alpha} - \sum_{j=-N}^{N} |v_j|^2 \frac{1}{E_j - \lambda}
\right ] \: . 
\label{eq:CPga}
\end{eqnarray}
The first factor of the product term is zero only 
at the  unperturbed eigenvalues $E_k$ of ${\cal H}^0$.
Therefore, for $\alpha \ne 0$ the solutions of Eq.~($\ref{eq:CPga}$)
must be given by the zeros of the second factor, i.e. by
the solutions of
$ \frac{1}{i\alpha} = \sum |v_k|^2 \frac{1}{E_j - \lambda}
 $. In the limit $\alpha \to 
\infty$,    there are  $2N$ solutions  lying at  real energies: 
   $\lambda_k\in (E_{k},E_{k-1})$  if $k>0$ and 
   $\lambda_k\in (E_{k},E_{k+1})$  if $k<0$
where $E_k$ is  eigenvalue of the unperturbed
Hamiltonian ${\cal H}^0$.
In the case of the picket fence with $E_k = k$ and equal coupling, $\lambda_k$
approaches $k \pm 1/2$.
Furthermore, we have exactly one complex solution at  ${E}_0=0$ and 
$\Gamma_0 \to \infty$ for $\alpha \to \infty$.
\\

Let us now discuss the behaviour of the system as a function of increasing
coupling strength $\alpha$.
>From Eq.~($\ref{eq:CPstrukt}$), we get
\begin{equation}
\frac{d\lambda}{d\alpha} Q'_N(\lambda) - i R_N(\lambda) - 
i\alpha \frac{d\lambda}{d\alpha} R'_N(\lambda) = 0 
\label{eq:CPga1}
\end{equation}
for the solutions of $P_N(\lambda) = 0$ and further
the differential equation 
\begin{equation}
\frac{d\lambda}{d\alpha}  =   \frac{i R_N(\lambda)}{Q'_N(\lambda) - 
i\alpha R'_N(\lambda)} 
\label{eq:CPga2}
\end{equation}
with the initial condition $\lambda_k(\alpha=0)=E_k$.
\\

For small $\alpha$, Eq.~($\ref{eq:CPga2}$) reads
\begin{equation}
\frac{d\lambda_k}{d\alpha}  \approx   \frac{i R_N(\lambda_k)}{Q'_N(\lambda_k)} 
 = - i |v_k|^2 \; .
\label{eq:CPga3}
\end{equation}
That means, the imaginary part of  
eigenvalue $\lambda_k$  of ${\cal H} $
increases, with increasing $\alpha$, proportional to 
$|v_k|^2$ while the real part of it remains unchanged, as long as $\alpha$ 
is small.\\

For large $\alpha$, we have $2N$ solutions whose imaginary part  is
small while the real part ${\cal E}_k$ is determined by
 $E_k < {\cal E}_k < E_{k-1}$ if $k>0$ and
 $E_k < {\cal E}_k < E_{k+1}$ if $k<0$,
since  $\lambda_k(\alpha\rightarrow\infty)={\cal E}_k$. 
 The relevant part
 of $R_N(\lambda)$ for the solutions of 
$P_N(\lambda) = 0$ is
 therefore $T_N(\lambda) = \prod_{k=1}^{2N} ({\cal E}_k - \lambda)$.
Inserting
\begin{equation}
\lambda_k(\alpha) = {\cal E}_k - i\frac{g_k}{\alpha} + O(\alpha^{-2}) \; .
\label{eq:ga}
\end{equation}
into 
\begin{equation}
0 = Q_N(\lambda) - i \alpha T_N(\lambda)
\end{equation}
leads in the two lowest orders in $1/\alpha$
to 
\begin{equation}
0 =  \prod_{j=-N}^{N} (E_j + i 
\frac{g_k}{\alpha}  - {\cal E}_k) 
- i \alpha \prod_1^{2N} ({\cal E}_j - {\cal E}_k + i g_k / \alpha) \; .
\label{eq:CPga5}
\end{equation}
The solution is
\begin{eqnarray}
g_k = - \frac{\prod_{j=-N}^{N} (E_j - {\cal E}_k)}{\prod_{j=1,j\ne k}^{2N} 
({\cal E}_j - {\cal E}_k)} > 0  \; .
\label{eq:CPga6}
\end{eqnarray}
Eq.~(\ref{eq:CPga5}) shows that for 
large coupling strengths, the decay widths of  $2N$ states decrease as
 $\frac{1}{\alpha}$ with increasing $\alpha$. This decrease is called 
resonance trapping.\\

 Besides these $2N$ solutions for large $\alpha$, we have a solution 
at $E_0 = 0$ and $\Gamma_0 \to
 \infty$ in the limit $N \to \infty$.
\\

In  sections 3 and 4, we will study in detail the properties of the 
characteristic polynomial 
 ($\ref{eq:CP}$) by means of  special cases. 
 \\

\subsection{The  eigenfunctions of a non-hermitian Hamilton operator}

Another  value characterizing the reorganization which
takes place in the open quantum system under
critical conditions, is the  mixing of 
 the wavefunctions of the resonance states \cite{ro91,ismuro}.
 The mixing caused by the coupling of all the
states to the common decay channels is related, in a natural manner, 
to the basic set of wavefunctions of the closed system,
\begin{eqnarray}
\Phi_i = \sum_{j=1}^M  a_{ij} \Phi^0_j 
\label{eq:entr1} 
\end{eqnarray}
where $\Phi_i $ are eigenfunctions of ${\cal H}$ and $ \Phi^0_j  $
are eigenfunctions of ${\cal H}^0$. The eigenfunctions $\Phi_i$
of the non-hermitian Hamiltonian ${\cal H}$ are bi-orthogonal. The right
and left eigenfunctions are defined by
\begin{eqnarray}
({\cal H} - \lambda_i) |\Phi_i^r \rangle & = & 0 
\nonumber \\
 \langle\Phi_i^l | ({\cal H} - \lambda_i)  & = & 0  
\label{eq:nonher1}
\end{eqnarray}
with the normalization
\begin{eqnarray} 
\langle \Phi_i^l | \Phi_j^r \rangle = \delta_{i,j} ,
\; \; \; 
\langle \Phi_i^r | \Phi_j^r \rangle \ne \delta_{i,j}
\; .
\label{eq:nonher2}
\end{eqnarray}
In our case  $|\Phi_i^l \rangle = (|\Phi_i^r 
\rangle)^T$ \cite{levrep}. In the following we will drop
 the indices $r$ and $l$
considering only the right eigenfunctions. Then the second relation of
Eq.~(\ref{eq:nonher2}) reads
\begin{eqnarray}
 \langle \Phi_i | \Phi_i \rangle \ge 1
\label{eq:nonher3}
\end{eqnarray}
and
$\langle \Phi_i | \Phi_j \rangle, \; i \ne j $, is a complex
 number, generally.
\\

A good numerical measure for the strength of mixing is
the number $N^p_i$ of principal components in the eigenfunction 
$\Phi_i$.
For its definition we are using the quantity
\begin{eqnarray}
b_{ij} = \frac{a_{ij}}{\sum_{l=1}^M |a_{il}|^2} \; .
\label{eq:coeffb}
\end{eqnarray}
Then, the number of principal components can be
calculated as
\begin{eqnarray}
N^p_i = \frac{1}{M\sum_{j=1}^M |b_{ij}|^4} \; .
\label{eq:nonher5}
\end{eqnarray}
The value of $N_i^p$ can be understood as a measure of (external)
 collectivity of
the resonance state $\Phi_i$. In the limiting case of equal mixing of the
state $i$ with  all states $j$,
$b_{ij} = 1 / \sqrt{M} \; \; \; \forall \; j$, we get $N^p_i = 1$
 (maximum external collectivity). In the opposite case (no external collectivity)
we have $b_{ij} = \delta_{i,j}$ and $N^p_i = 1 / M$. Generally,
 $1 / M \le N_i^p \le 1$. 
\\ 

 Further, 
 we introduce the value 
\begin{eqnarray}
B = \frac{1}{M} \sum_{i=1}^M \langle \Phi_i | \Phi_i \rangle
\ge 1  
\label{eq:nonher4}
\end{eqnarray}
 which characterizes the degree of non-Hermiticity
of ${\cal H}$ according to 
Eq.~(\ref{eq:nonher3}). It is a function of $\alpha$ and  $B = 1$ if 
${\cal H}$ is  hermitian. 
\\

\section{The ideal picket-fence distribution}

Let us consider first the simple case of a picket-fence distribution of $M=2N+1$
levels which are all coupled with the same strength (''ideal picket-fence
distribution'') to the continuum
consisting of one decay channel ($K=1)$. The advantage of this simple model is
that analytical studies can be performed.
\\

\subsection{Analytical study for the limiting case $N\to \infty$}

Suppose
 $E_k = k$ und $|v_k| = 1 \;\;\; \forall k$. Then
Eq.~($\ref{eq:CP}$) reads
\begin{eqnarray}
P_N(\lambda)
&  \equiv & Q_N(\lambda) - i \alpha \cdot R_N(\lambda)
\nonumber \\
 & = & \prod_{k=-N}^{N} (k - \lambda) - 
i\alpha \cdot \sum_{k=-N}^{N}
 \prod_{j = -N, j\ne k}^{N} (j - \lambda)
\label{eq:CPpf}
\end{eqnarray}
and the relation ($\ref{eq:PFzus}$) holds.
In order to consider the limit $N\rightarrow\infty$, we divide   
 $Q_N$ by a convergence ensuring factor,
\begin{eqnarray}
\lim_{N\to\infty}\frac{Q_N(\lambda)}{-\prod_{k=1}^{N} -(k)^2}  = 
 \lim_{N\to\infty} 
\lambda\cdot\prod_{k=1}^{N} (1-\frac{\lambda}{k})(1+\frac{\lambda}{k})
= \lambda\cdot\prod_{k=1}^{N} (1-(\frac{\lambda}{k})^2) =
 \frac{\sin(\pi\lambda)}{\pi}
\label{eq:CPpf1}
\end{eqnarray}
Then the characteristic polynomial reads
\begin{equation}
P(\lambda) = \sin(\pi\lambda) + i\pi\alpha \cos(\pi\lambda) = 0
\; .
\label{eq:CPtrig}
\end{equation}
Denoting the complex eigenvalue of ${\cal H}$ by $\lambda = {\cal E} - i \frac{\Gamma}{2}$
and splitting Eq.~(\ref{eq:CPtrig}) into its real and imaginary part
we get (for real $\alpha$):
\begin{eqnarray}
\cos(\pi {\cal E}) \left [ e^{\pi\Gamma} (1 -\pi\alpha) - (1 + \alpha\pi)\right ] & =
& 0 
\nonumber \\
\sin(\pi {\cal E}) \left [ e^{\pi\Gamma} (1 -\pi\alpha) + (1 + \alpha\pi) \right ] & = & 0
\label{eq:CPri}
\end{eqnarray}
Since the two functions $\cos(x)$ and $\sin(x)$ never vanish for the same argument
$x$, we have to consider two different cases:
\begin{enumerate}
\item
 $\sin(\pi {\cal E})=0 \Rightarrow {\cal E}=n\in Z$ and
\begin{equation}
e^{\pi\Gamma}=\frac{1 + \pi\alpha}{1 - \pi\alpha}
\label{eq:CPnum1}
\end{equation}
has a real solution $\Gamma$ for $\alpha < \frac{1}{\pi}$ only.
For small $\alpha$, we have therefore 
\begin{equation}
\Gamma = \frac{1}{\pi} \; \ln\left ( \frac{1 + \pi\alpha}{1 - \pi\alpha}
\right ) 
\label{eq:CPite1}
\end{equation}
and $\; \Gamma \to 
-\frac{1}{\pi}\; \ln \; \varepsilon$ ~for $\alpha 
= \frac{1}{\pi} (1 - \varepsilon)$ and $\varepsilon \to 0$.

\item
 $\cos(\pi {\cal E}) = 0 \Rightarrow {\cal E}=n+\frac{1}{2}$ for
$n\in Z$ and
\begin{equation}
e^{\pi\Gamma}=\frac{\pi\alpha + 1}{\pi\alpha - 1} \; .
\label{eq:CPnum2}
\end{equation}
The last equation can be fulfilled only for  $\alpha >\frac{1}{\pi}$.
For large  $\alpha$ it is therefore 
\begin{equation}
 \Gamma=\frac{1}{\pi} \ln\left (\frac{\pi\alpha + 1}
{\pi\alpha - 1} \right ) 
\label{eq:CPite2}
\end{equation}
and $\; \Gamma \to 
-\frac{1}{\pi}\; \ln \; \varepsilon$ ~for $\alpha =
 \frac{1}{\pi} (1+\varepsilon)$ and $\varepsilon \to 0$.
\\
\end{enumerate}

As a result: The widths of all the states increase up to infinity 
as a function of increasing $\alpha$.
The singularity at the critical point $\alpha_{{\rm crit}}$ is determined by
$\ln(\varepsilon)$.  It is {\it logarithmic}.
\\

Further, the energetical positions of the states remain unchanged
 at the unperturbed energies 
$E_k=k$ of the system (eigenvalues of ${\cal H}_0$)
up to $\alpha\rightarrow\frac{1}{\pi}$.
At $\alpha_{{\rm crit}} = \frac{1}{\pi}$, the real part ${\cal E}_k$ of  $2 N$  
eigenvalues (all $k\ne 0$) of ${\cal H}$ jumps from $k$ to 
     $k-\frac{1}{2}$ if $k>0$ and from $k$ to  $k+\frac{1}{2}$ 
if $k<0$, respectively. As a function of further increasing $\alpha$,
 the imaginary part of the eigenvalues of the $2 N$ resonance states 
 (all $k$ but $k=0$) decreases first as  $\ln(\alpha)$ while
it approaches zero as $\frac{1}{\alpha}$ for  $\alpha \to \infty$ 
according to Eq.~(\ref{eq:CPga}).\\

In order to study the behaviour of the state in the centre of the spectrum at
the energy ${\cal E}_0 = 0$ we 
consider
only the highest-order terms of $\lambda$
 in Eq.~(\ref{eq:CP}): 
\begin{equation}
\lambda^{2N+1} + i\alpha (2N+1) \lambda^{2N} = 0 \; .
\label{eq:CPnum3}
\end{equation}
For large $\alpha$ ($\alpha \gg \frac{1}{\pi}$),
 the state corresponding to the solution $\lambda = - i
\alpha (2N+1)$ ~lies at ${\cal E}=0$ and its width increases linear with $\alpha$. 
 \\

Summarizing the results, we state the following:
In spite of the fact, that the coupling parameter $\alpha$ enters  the equations
(\ref{eq:ham}) and (\ref{eq:CPstrukt})
linearly, the imaginary parts of the complex eigenvalues
show a singularity at the finite value  $\alpha=\frac{1}{\pi}$. 
For larger couplings, a clear separation of the time scales with respect to the decay widths
of the resonance states occurs. This happens also in the case of an infinitely
extended spectrum. This is not a local effect of a {\it locally} broad
resonance in a restricted energy region, but it is produced by the whole system
in a {\it collective} manner.  {\it All} basic states, independent of 
their energy position, 
act cooperatively.\\

\subsection{Widths at the critical point for finite $N$: analytical study}

The sum of the widths of all states is, in our simple example with equal 
coupling strengths, given by
\begin{eqnarray}
\Im \{Tr({\cal H}) \} = \sum_{j} \frac{\Gamma_j}{2} = \alpha (2N+1) \; .
\label{eq:summenf}
\end{eqnarray}
It is $Tr({\cal H}) = {\rm const}  \; (\alpha)$. Thus,
 $\Im \{Tr({\cal H}) \}$
should be  a smooth function of $\alpha$  not only far from the  
critical point 
but also near to it
in spite of the divergence of the widths for $N\to \infty$
at $\alpha = \frac{1}{\pi}$ (see subsection 3.1).
In the following, we will proof this statement. 
\\

First, let us consider the eigenvalues of ${\cal H}$  for finite $N$.
In this case, we have
\begin{eqnarray}
P_N(\lambda) = \prod_{k=-N}^{N} (k-\lambda) \left [ 1 -
 i\alpha \sum_{j=-N}^{N} \frac{1}{j-\lambda}
\right ] = 0
\label{eq:Pend}
\end{eqnarray}
 instead of the simple Eq.~(\ref{eq:CPtrig}) holding for $N\to \infty$.
As discussed in relation with  Eq.~($\ref{eq:CPga}$),
 the solutions of $P_N(\lambda) = 0$ follow from 
 $1 - i\alpha \sum_{j=-N}^{N} \frac{1}{j-\lambda} = 0$.
Here, we are interested in the difference  between the solutions obtained
 for finite $N$ and those for $N \to \infty$
\\

It holds
\begin{eqnarray}
0 = 1 - i \alpha \sum_{k=-N}^{N} \frac{1}{k-\lambda} =
 1+ i\alpha\pi \cot(\pi\lambda) +
2 i \alpha\lambda \sum_{k=N+1}^{\infty} \frac{1}{k^2 - \lambda^2}
\label{eq:endSum}
\end{eqnarray}
where the correction term is given by
\begin{eqnarray}
2 \lambda \sum_{k=N+1}^{\infty} \frac{1}{k^2-\lambda^2} & \approx & 2 
\lambda \int_{N+1/2}^{\infty} 
\frac{dx}{x^2-\lambda^2} =
 2 \int_{\frac{N}{\lambda}}^{\infty} \frac{dy}{y^2-1} 
\nonumber\\
& = & \left [ \ln\frac{y-1}{y+1} \right ]_{\frac{N}{\lambda}}^{\infty}  = 
\left [ \ln\frac{1-1/y}{1+1/y} \right ]_{\frac{N}{\lambda}}^{\infty} 
\nonumber\\
& = &  \left [ \ln(1-\frac{2}{y} + O(y^2))
 \right ]_{\frac{N}{\lambda}}^{\infty}
\approx \left [ -\frac{2}{y} \right ]_{\frac{N}{\lambda}}^{\infty} = 
\frac{2 \lambda}{N}
\end{eqnarray}
under the assumption
 $1/y=\lambda /N \ll 1$. This condition is fulfilled, to a
good approximation, 
in the centre of the spectrum. Splitting
 Eq.~(\ref{eq:endSum})
into its real and imaginary parts (with $\lambda = {\cal E} - \frac{i}{2}\Gamma$), one arrives at
\begin{eqnarray}
0 = 1 - \alpha \pi \frac{\sinh(\pi\Gamma)}{\cosh(\pi\Gamma) -
 \cos(2\pi {\cal E})} + \alpha \frac{\Gamma}{N}
\end{eqnarray}
for the real part. 
Here the identity  
\begin{eqnarray}
\cot(x+iy)=\frac{\sinh(2x) -i \sin(2y)}{\cosh(2y) - \cos(2x)}
\end{eqnarray}
is used. The equation for the imaginary part 
 (${\cal E}\ne 0$) reads
\begin{eqnarray}
0 & = & \alpha \pi \frac{\sin(2\pi {\cal E})}{\cosh(\pi\Gamma) -
 \cos(2\pi {\cal E})} + \frac{2\alpha {\cal E}}{N} 
\end{eqnarray}
from which we get
\begin{eqnarray}
0 = \sin(2\pi {\cal E}) + \frac{2 {\cal E}}{\pi N} \cosh(\pi\Gamma) -
  \frac{2 {\cal E}}{\pi N} \cos(2\pi {\cal E}). 
\end{eqnarray}
An estimation for the upper limit of the widths $\Gamma$ 
of the states at $\alpha = \frac{1}{\pi}$
  leads to
\begin{eqnarray}
\pi\Gamma = {\rm arcosh} \left [ \cos(2\pi {\cal E}) - \frac{N\pi}{2 {\cal E}}
\sin(2\pi {\cal E}) 
\right ]
\approx \ln\frac{N\pi}{|{\cal E}|} \; .
\label{eq:huell}
\end{eqnarray}
Here, we have used $N / |{\cal E}| \gg 1$
which is fulfilled only in the centre of the spectrum.
Thus,
\begin{eqnarray}
\frac{\Gamma}{2}(\alpha=\alpha_{{\rm crit}}) \; \leq \;
 \frac{1}{2\pi} \; \ln \; \frac{N\pi}{|{\cal E}|}
\label{eq:oSchr} 
\end{eqnarray}
which holds for every one
of the  $2N$
states (for all $k$ but $k=0$) at the critical point. It means,
 $\Gamma \leq \ln \; N$
for $\alpha \to \frac{1}{\pi}$ for {\it all} $N$.\\

Using    
Eq.~(\ref{eq:oSchr}), one gets the following estimation for the trace of the imaginary part of 
$H_{QQ}^{\rm eff}$
at $\alpha = \frac{1}{\pi}$:
\begin{eqnarray}
\sum_{j=-N}^{N} \frac{\Gamma_j}{2} & \approx & 2 \int_{0}^{N} dE
 \frac{1}{2\pi} \ln \frac{N\pi}{E} 
= -N \int_{0}^{1/ \pi} \ln(x) dx = -N \left [ x \ln(x) -x \right ]_{0}^{1/ \pi}
\nonumber \\
& & = \frac{N}{\pi} (1 - \ln\frac{1}{\pi}) = \frac{N}{\pi} (1 + \ln(\pi))
 \approx \frac{2 N}{\pi}
\approx \frac{2N+1}{\pi} \; .
\label{eq:summenfe}\end{eqnarray}
   
The comparison of
Eqs.~(\ref{eq:summenf}) and (\ref{eq:summenfe}) 
shows that
Eq.~(\ref{eq:summenf})
 holds also at the critical point.
This means, the singularity of the decay widths 
$\Gamma$ at the critical point
occurs such
that the sum rule $\sum_i \Gamma_i = {\rm const}  (\alpha)$ is fullfilled also for
$\alpha\to\alpha_{{\rm crit}}$ and $N \to \infty$. 
\\

At the critical point,
the width $\Gamma_0$ of the state in the centre of the spectrum
can be estimated
in leading order in $N$  by integrating  
Eqs.~(\ref{eq:huell})
over the interval $(-1/2, 1/2)$:
\begin{eqnarray}
\frac{\Gamma_0}{2} (\alpha = \alpha_{{\rm crit}}) =
\frac{1}{2 \pi} \int_{-1/2}^{1/2} \ln  \bigg(\frac{N \pi}{|E|}\bigg) dE
= \frac{1}{2 \pi} (1+\ln  (2 \pi N)) \; .
\label{eq:estimation}
\end{eqnarray}
Thus, the width of the broadest state at the critical point is small in comparison
to the total length $2N$ of the spectrum. 
\\

\subsection{Numerical illustration}

In the following, we illustrate  the behaviour of the decay widths 
by results of numerical studies
for different $\alpha $  and for some finite values of $N$ between
 50  and 5000.\\

Splitting the sum for finite $N$ in 
Eq.~(\ref{eq:endSum}) into its real and imaginary part,
one gets
\begin{eqnarray}
0 = \sum_{k=-N}^{N} \frac{k-{\cal E}}{(k-{\cal E})^2 + (\Gamma/2)^2} 
\label{eq:eSimag}
\end{eqnarray}
and
\begin{eqnarray}
1 = \alpha \sum_{k=-N}^{N} \frac{\Gamma/2}{(k-{\cal E})^2 + (\Gamma/2)^2} \; .
\label{eq:eSreal}
\end{eqnarray}
The first equation describes the trajectories of the eigenvalues
$\lambda = {\cal E} - \frac{i}{2} \Gamma$ of
${\cal H}$ in the complex plane while the second one contains their 
parametrization with $\alpha$. We calculated the sum in
  Eq.~(\ref{eq:eSimag})
for different $N$ and fixed values of $\Gamma$
and traced their solutions
as a function of ${\cal E}$.\\

In Fig.~1.a, 
 the results of the calculations  for different $N$
($N=500, 1000, 5000$) and fixed $\Gamma = 1$ are   shown
as a function of ${\cal E}$.
Due to the denominator of the sum, every resonance state $k$ with
$\Gamma_k \ge 1$ has two solutions  while there are no solutions when 
$\Gamma_k < 1$. The number of solutions  and thus the number 
of resonance states with $\Gamma_k > 1$
depends on the average slope
by which the sum approaches the value $0$ as a function of ${\cal E}$. \\

The result is as follows: 
For a fixed value of $\Gamma = 1$
 there are the more solutions of  Eq.~(\ref{eq:eSimag}), 
the larger $N$.
In the limit $N \to \infty$,    Eq.~(\ref{eq:eSimag}) can be fulfilled 
for all resonances, i.e. all resonance states have widths larger than 
an arbitrarily chosen finite value. This confirms the analytical result for
an infinite number of states, where we have shown that the widths of {\it all} 
states diverge at $\alpha=\alpha_{{\rm crit}}$.\\

In Fig.~1.b, the results of calculations are shown with
a fixed number $N=1000$ and different values of $\Gamma$ 
($\Gamma = 0.5, 0.75, 1.0$).
In this case, the average slope of the different curves is the same but the
amplitude of the oscillations varies.  The larger $\Gamma$ the smaller the 
amplitude is. That means, in the case of finite $N$, 
the number of resonance states having $\Gamma_k \ge \Gamma$
is the smaller the larger $\Gamma$.\\

As a result, we state the following:
 When $N$ is a finite number,
 only a limited number of resonance states has widths $\Gamma_k \ge \Gamma$ where 
$\Gamma$ is an arbitrarily chosen  finite value. 
Further, 
 the larger $N$, the larger  the number  of resonance states
with  $\Gamma_k \ge \Gamma$. 
On the other hand, the larger $\Gamma$,
the smaller the number of resonance states with $\Gamma_k \ge \Gamma$.
Thus we have {\it two processes compensating  each other} 
which ensures, that  (\ref{eq:summenf}) holds also in the limit $\alpha \to 
\frac{1}{\pi}$ and $N \to \infty$.
\\

In Fig.~2.a, we illustrate the motion of the eigenvalues in the complex plane 
for $0.01 < \alpha < 2$ in steps of $0.01$ for positive energy ${\cal E}$
 (the part for negative energy is 
symmetric to that for positive energy). 
For each resonance state its eigenvalue follows a certain trajectory
with increasing $\alpha$.
For the lowest values of $\alpha$, all
eigenvalues are near to $E_k = k$ and $\Gamma_k / 2 = \alpha$. The full
line gives the estimation for the upper limit of $\Gamma /2$ at $\alpha =
\alpha_{{\rm crit}}$ according to  Eq.~(\ref{eq:oSchr}).
 The estimation is good in the centre of the spectrum. The deviations 
at large energies are pure boundary effects. The differences between the
different eigenvalue trajectories arise from the finite value of $N$. In the limit
of $N \to \infty$ all eigenvalues acquire 
the same behaviour because of the
discrete translational symmetry on the real energy axis.\\

We show in Fig.~2.b the behaviour of all $\Gamma_k / 2$ as a function of
$\alpha$ for $N=50$. At the critical value $\alpha_{{\rm crit}} = 1 / \pi$ 
 (indicated by a vertical solid line) the width of the collective  resonance 
state $k=0$ separates from the widths of the other ones
and increases linearly. The slope of 
$\Gamma_0 (\alpha) / 2$ is equal to $2N+1$ over almost the whole range of
$\alpha > \alpha_{{\rm crit}}$ according to 
 Eq.~(\ref{eq:CPnum3}).
The larger slope of $\Gamma_0 (\alpha)$ close to $\alpha_{{\rm crit}}$ is a
boundary effect and disappears in the limit of $N\to\infty$.
 \\

Fig.~2.c shows $N_0^p$ as a function of $\alpha$ for two different values of
$N$. The curves show a sudden rise at $\alpha_{{\rm crit}}$
and saturate rapidly to 1. The inlet gives a magnification
of the curve around the critical
point  which is marked by a vertical solid line. The larger $N$, the sharper are
 the changes in the slope of 
$N_0^p(\alpha)$.
This is a clear numerical indication of the cooperative effect 
acting over the {\it total} length of the spectrum. 
In spite of the fact, that the width $\Gamma_0$ of the fast decaying state at
 $\alpha\approx\alpha_{{\rm crit}}$
is of the order $\ln(M)$, eq. (\ref{eq:estimation}),
its wavefunction carries   contributions from basis states
which are lying  far away from the centre of the spectrum
and are {\it not} overlapped by it. 
These contributions over large energy scales 
are achieved via the "chain" of overlapping
neighbouring  
resonances. As a result, we
observe a  "macroscopic" order
 over the whole energy scale of  the system being much larger than 
$\Gamma_0$. 
\\

The curve of $N_0^p$ does, of course, not jump immediately
 to $100\%$ at the critical point. The main reason
is the finite number of states taken into account in the calculations.
The widths of the resonances at the border of the spectrum are in general
smaller than those of the resonances inside the spectrum. 
Therefore the chain of neighbouring 
overlapping states
is interrupted at energies close to the edges.\\

For all the trapped states the corresponding quantity $N_k^p$
always remains in the order of $1/M$. The wavefunctions 
of these states are 
mixed only with those of their next neighbours.
\\
 
All the numerical results show that although the phase transition
appears mathematically for
 $N \to \infty$ only,
the characteristic features of it can be already seen at {\it 
 comparably} small values of $M$. 
\\

\subsection{Picket-fence level distribution with disturbed translation 
invariance}

Let us break the translation invariance of the picket-fence model by giving
another coupling strength to the state  in the centre of the system. 
Suppose: $E_k=k$ und $|v_k|=1 \;\;\; \forall k-\{0\}$ and $v_0=1+D$.
In this case, the characteristic polynomial
($\ref{eq:CP}$) reads
\begin{equation}
P_N(\lambda) = \prod_{k=-N}^{N} (k-\lambda) - 
i\alpha\sum_{k=-N}^{N}\prod_{j\ne k}
(j-\lambda) - i\alpha D\prod_{k=1}^{N} (k-\lambda)(k+\lambda) \; .
\label{eq:CPoTI}
\end{equation}
Dividing by a convergence ensuring
factor and
identifying the resulting terms with the product representation of
 $\sin(x)$ and $\cos(x)$, respectively,
we get
\begin{equation}
P(\lambda) = \frac{\sin(\pi\lambda)}{\pi} + i\alpha \cos(\pi\lambda) +
 i\alpha D 
\frac{\sin(\pi\lambda)}{\pi\lambda} = 0
 \end{equation}
in the limiting case $N\rightarrow\infty$. In order to study the behaviour of
 the state in the centre  (${\cal E}=0$) we write 
 $\lambda=-i\mu$ and get
\begin{equation}
\alpha  =  \frac{1}{\pi} \left [ \frac{1}{\frac{D}{\pi\mu} + \coth(\pi\mu)}
 \right]  \; .
\label{eq:CPoTItri}
\end{equation}
According to this equation, 
 $\alpha \to \frac{1}{\pi}$ for   $\mu \rightarrow\infty$.
The redistribution of the system takes place  
at the same finite value of $\alpha_{{\rm crit}} = 1 / \pi$ as in the case
of constant coupling.
\\

We investigate now the behaviour of the system at the critical point,
i.e. the type of the singularity.
Suppose
 $\varepsilon = 1 - \pi\alpha$ and 
$\pi\mu = c \varepsilon^{-s}$ with $s\in R$.
Using the relation
$\coth(\pi\mu)\rightarrow 1$ for large $\mu$,
we get from
 (\ref{eq:CPoTItri})
\begin{equation}
1 = (1 - \varepsilon)(1 + \frac{D}{c}\varepsilon^{+s}) \; .
\label{eq:Salg}
\end{equation}
In leading order of the singular part of  $\pi\mu(\varepsilon)$, 
this equation is solved
by $s=1$ and $c=D(1-\varepsilon)\approx D$.
That means, we have an 
 {\it algebraic} singularity:  
$(\pi\mu = \frac{D}{\varepsilon})$.
The system approaches the singularity quicker than in the case of a picket
fence with translation invariance. 
\\

In the following we like to discuss formula (\ref{eq:CPoTItri})
in detail with the help of a numerical illustration (Figs.~3 and 4).
It turned out, that the case with a discriminated state at 
${\cal E}=0$ is  more complicated.
Therefore we have drawn the movement of the complex eigenvalues
 $\lambda$ in the centre of the spectrum
for $D=-0.5$ (Fig.~3.a). We use
$M=2N+1=101$ and $301$, respectively.  \\

At $\alpha=\alpha_{{\rm crit}}=1/\pi$ two broad modes arise
at the flanks of the spectrum at $|{\cal E}|\approx 4.5$ and $7.5$
for $N=50$ and 150, respectively. The larger the number of resonance
states is,
the larger is their distance from the centre. With increasing $\alpha$ the poles 
of the two resonance states at positive and negative energy
approach each other in their real part and  collide at ${\cal E}=0$ at a certain
 value $\alpha=\alpha_{c1}$. 
At $\alpha > \alpha _{c1}$, the resonance states remain 
at ${\cal E}=0$, one with further increasing,
the other  one with decreasing $\Gamma$.
The more resonance states the spectrum contains
the smaller is $\alpha_{c1}$,
as one can see in  Fig.~3.b. Here the imaginary parts $\Gamma_k / 2$
 of  the eigenvalues $\lambda_k$  shown in  Fig.~3.a are drawn
as a function of the coupling strength $\alpha$.
 The collision point shifts to the critical point
of the spectrum ($\alpha_{c1}\rightarrow\alpha_{{\rm crit}}$) 
if the number $M$ of resonance states is enlarged, in spite of the fact that
 their distance from one another  at $\alpha_{c1}$ is larger when $M$ is
larger.  The two values of $\alpha_{c1}$ corresponding to
 $M=101$ and $M=301$, are indicated by
vertical dashed lines. The values of $\alpha_{\rm crit}$ and $\alpha_{c2}$ are
marked by vertical solid lines.
In the limit $M\rightarrow\infty$ the two broad poles
appear  at ${\cal E}\to \infty$. In this limit, 
$\alpha_{c1} \to \alpha_{{\rm crit}}$.  
At this point of $\alpha$, the poles jump  to ${\cal E}=0$. \\

Between $\alpha=\alpha_{c1}$
and the finite value $\alpha=\alpha_{c2}$, there exist three  
resonance states at ${\cal E}=0$:
the two broader poles appearing at $\alpha_{c1}$  and the original one
 which is discriminated
by the external coupling by $D$.
The collective mode is one of the two resonance states
arising  from the phase transition at $\alpha = \alpha_{c1}$. 
Its imaginary part
increases with further increasing $\alpha$.
The other broad pole decreases in $\Gamma$ with increasing $\alpha$.
Its collision with the discriminated resonance state
at $\alpha=\alpha_{c2}$  shifts both states  away from ${\cal E}=0$
and the imaginary part of both eigenvalues decreases. 
Contrary to the value of $\alpha_{c1}$ which approaches $\alpha_{{\rm crit}}$
with $M\rightarrow\infty$
the value of $\alpha_{c2}$ remains almost constant as a function of $M$.
For $M\to\infty$, the value of $\alpha_{c2}$ remains larger
than  $\alpha_{{\rm crit}}$.
As we will see below, it mainly depends on $D$.
The poles of the trapped
states approach the values $n+1/2$ (with $n\in Z$)
 if $\alpha\rightarrow\infty$.\\

Figs.~4.a. and ~4.b. show the graphs of $\pi\mu \coth(\pi \mu) + D$ for
different $D$ $(D=-0.5,0,0.5)$ and  $\mu / \alpha$ for
 several values of $\alpha$ as a function of $\mu$.
The points of intersection are the solutions of equation (\ref{eq:CPoTItri})
 derived under the assumption of an infinite number of states.\\

In Fig.~4.a. the coupling parameter is set to  $\alpha=0.1<\alpha_{{\rm crit}}$ and 
$\alpha= 1 / \pi =\alpha_{{\rm crit}}$, respectively.
For the value $\alpha < \alpha_{{\rm crit}} $ there exists only one point 
of intersection with each of
the curves of $\coth$, lying at small values of $\mu$.
Also the state at ${\cal E}=0$ has a comparably small width in the
 undercritical regime of $\alpha$
and the value of $\mu$ increases with increasing $D$.
In the parameter range $\alpha<\alpha_{{\rm crit}}$ no broad mode
 is separated from the other ones. \\

At the critical point $\alpha=\alpha_{{\rm crit}}$, where the phase
 transition takes place, the linear curve
$\mu / \alpha$ is tangential to the $\pi\mu \coth(\pi \mu) + D$
 for $D=0$ and is parallel to
this function for $D=-0.5,0.5$ (lower and upper  thick full line, respectively).
 So each of these curves has a point
 of intersection with $\mu / \alpha$ at $\mu=\infty$. 
For the case  $D= - 0.5$, the intersection at $\mu=\infty$ contains 
two solutions for the
two broad modes, arising at the borders (at ${\cal E}=\pm\infty$) of the spectrum and
 colliding at ${\cal E}=0$.
Additionally, there is another intersection  with $\mu/\alpha$
at a small value of $\mu$ which arises from the discriminated state at 
$E_0=0$. 
 \\

In Fig.~4.b, we see the same curves  $\pi\mu \coth(\pi \mu) + D$ as
 in Fig.~4.a together with
$\mu/\alpha$ for different values of $\alpha > \alpha_{{\rm crit}}$:
$\alpha < \alpha_{c2}$,
$\alpha\approx\alpha_{c2}$ and $\alpha>\alpha_{c2}$.
In all cases, the curves have intersections at $\mu=\infty$. This
means, 
for all values of $D$ 
a mode exists  at ${\cal E}=0$ with an infinitely 
large width
if $\alpha\ge\alpha_{{\rm crit}}$.
The curve $\pi\mu \; \coth(\pi \mu) -0.5$ shows three intersections with 
$\mu/\alpha$ in the range
$\alpha_{{\rm crit}}<\alpha<\alpha_{c2}$. 
The two intersections at smaller $\mu$ 
approach  each other if $\alpha\rightarrow\alpha_{c2}$.
The value of $\alpha_{c2}$ depends obviously on the value of $D$.
In dependence on the negative shift of  $\coth(\pi \mu)$ the slope of
the tangent  changes and therefore the value of $\alpha_{c2}$.
At larger values of the coupling $(\alpha>\alpha_{c2})$ only one solution remains at ${\cal E}=0$ with
 $\mu=\infty$.
\\

Finally we have investigated the behaviour of the number
$N^p_0$ of principal
components 
of the broad resonance as a function of $\alpha$
 for   $D=\pm 0.5$ and $M=101$ states.
The results are drawn in Fig.~5 together with the former one
 for $D=0$.
For all cases the sudden rise of $N^p_0$ at $\alpha\approx\alpha_{{\rm crit}}$ can be seen.
At the critical point the collective  state
is created, in all cases, by almost all basis states distributed over the whole spectrum.
This collective behaviour of a
phase transition is well pronounced also for
$D\ne 0$. The  curve for $D=0.5$ is, however, much smoother 
than the other ones.
 It rises up not as
quickly as the other ones and does not
 approach the maximum value
 $100\%$ in the  range of $\alpha$ shown in the figure. Nevertheless,
 the characteristics of a
 phase transition can be seen also in this numerical study:
Comparing the results for $M=301$ states
in the case of $D=0.5$
 with those for $M=101$ states  we see the following tendency.
By increasing the number of states, $N^p_0$ for $D=0.5$ comes closer
to the curves $N^p_0$ for  $D=1$ and $-0.5$.
 This behaviour is a hint to  a phase transition 
(in the next section, Fig.~7.a, we show an example in which
  a phase transition does not occur and
the  results as a function of an increasing number $M$ of 
states do not show such a tendency).
\\

\subsection{Phase transition}

Summarizing the results of our study with the ideal picket fence,
 we state the following.

\begin{itemize}  
\item[--] 
For an infinite number of states the imaginary parts of the complex eigenvalues
of the effective Hamiltonian (\ref{eq:ham}) show a singularity at the finite value 
$\alpha = \alpha_{{\rm crit}} = 1 / \pi$. 
A bifurcation in the widths appears: the width $\Gamma_{i=0}$  of the
state in the centre of the spectrum increases  with  further increasing
$\alpha$ while the widths of
all the other states start to decrease.
All states but the state at $E=0$ are shifted by 1/2 at 
$\alpha = 
\alpha_{{\rm crit}}$.
For finite $M$ the width $\Gamma_{i=0}$ increases linearly with
 $\alpha > \alpha_{{\rm crit}}$
with a slope given by the number $M$ of states included in the spectrum.

\item[--] 
The state in the centre of the spectrum is a collective one
in a global sense. 
The numerical studies for finite systems  show that it contains 
components of almost {\it all} basic states of the system,  also
of those which are {\it not} overlapped by it. 
This is expressed by  
 the number $N_{i=0}^p$ of principal components of the state in the centre of
the spectrum which
grows,  at $\alpha = \alpha_{{\rm crit}}$, suddenly from its minimum 
value to $100\%$ (maximum mixing).

\item[--] 
 At $\alpha = \alpha_{{\rm crit}}$, the width $\Gamma_{i=0}$ of the state in the
centre of the spectrum is much smaller than the
extension of the spectrum. Therefore,
the system does   {\it not} create locally a collective state which
 traps, with  increasing
$\alpha$, further resonance states overlapped by it.

\item[--]
At $\alpha = \alpha_{{\rm crit}}$, the system suffers a change in its structure:
one of the resonance states aligns with the decay channel. Its 
wavefunction collects all the corresponding  
 components from the wavefunctions of all the other states
(which appear with the same weight in all basic wavefunctions 
because of the symmetry of the problem).
Therefore, its width $\Gamma_{i=0}$ increases with further 
increasing $\alpha$ while
the widths of all the other states decrease.

\end{itemize}

Generally,
the behaviour of the order parameter of a system 
as a function of a control parameter characterizes the type
of the phase transition.
In case of a first order phase transition the order parameter
shows a jump at
a certain finite value of the control parameter 
$\alpha=\alpha_{\rm crit}$.
If a higher order phase transition is present, the corresponding
derivative of the order parameter jumps at 
$\alpha_{\rm crit}$. More precicely,
its $(n-1)$-th derivative jumps in case of an $n$-th order phase transition.\\

In our system, the value $\Gamma_0 / M$ can be understood
 as the order parameter. It is the width of the collective state normalized
according to the number of resonance states  contributing to its
 formation. 
This value increases linearly as a function of $\alpha$  with the slope $1/M$ for
$\alpha < \alpha_{{\rm crit}}$ and with 
the slope one if $\alpha > \alpha_{{\rm crit}}$. 
So, 
the first derivative of $\Gamma_0 / M$  jumps at $\alpha=\alpha_{\rm crit}$.
Therefore we conclude, that the formation of 
a globally collective resonance is a {\it second order} phase transition.
We will prove in the next sections, that this behaviour
is universal for all systems showing a collective reorganisation.
\\

In sections 4.5 and 5, we will see that the phase transition is accompanied by an
essential deviation of the value $B$
(Eq.~($\ref{eq:nonher4}$)) from 1 in the neighbourhood of  
 $\alpha_{{\rm crit}}$. 
This means the bi-orthogonality of the function
system plays an important role in the reordering process.
\\

\section{More realistic systems}

In the previous 
section, we investigated the properties of the 
translation invariant
picket-fence model as a function of the coupling parameter 
$\alpha$ analytically as well as numerically. The system suffers a 
second-order phase transition at $\alpha = \frac{1}{\pi}$ which we studied in
detail. We  consider now the behaviour of some more realistic  systems
as a function of the parameter $\alpha$ in order to  answer 
the question whether the results obtained have a general meaning.
In detail, energy dependencies and fluctuations within the spectrum will 
be considered in the following.\\

\subsection{System with unequally distributed  levels and equal coupling 
strength}

We investigate now the behaviour of  systems when the level density 
is not constant but changes as a function of energy. 
Suppose:
$E_k = {\rm sign}(k)\; k^2$ und $v_k = 1 \;\;\; \forall k$.
This means, the level density is assumed to decrease linearly with
 $k$ and to approach the value zero for  $k\rightarrow\infty$.
\\

In this case, the characteristic polynomial reads 
\begin{equation}
P_N (\lambda) = \prod_{k=-N}^{N} \Big({\rm sign}(k) \; k^2 - \lambda\Big) - 
i\alpha \sum_{k=-N}^{N} \;
\prod_{j=-N; j\ne k}^N \Big({\rm sign} (j) \; j^2 - \lambda\Big) 
\label{eq:CPk2}
\end{equation}
which can be rewritten as
\begin{equation}
P_N(\lambda) = Q_N(\lambda) + i\alpha \cdot \frac{d}{d\lambda} Q_N(\lambda)
\end{equation}
where $Q_N(\lambda) =  \prod_{k=-N}^{N}  \Big({\rm sign}(k)\; k^2 - \lambda\Big)$.
Dividing $Q_N(\lambda)$ by the convergence 
ensuring factor $F = -\prod_{k=1}^N -(k)^4$ we get
\begin{equation}
\frac{Q_N(\lambda)}{F} = \lambda \prod_{k=1}^N
 (1-\frac{\lambda}{k^2})(1+\frac{\lambda}{k^2}) \; .
\end{equation}
In the limit $N \to \infty$, this expression is  
\begin{eqnarray}
\frac{Q_N(\lambda)}{F}  & =  & \frac{1}{\pi^2} \sin(\pi\sqrt{\lambda})
 \sinh(\pi\sqrt{\lambda})
 =  \frac{-i}{\pi^2} \sin(\pi\sqrt{\lambda}) \sin(i\pi\sqrt{\lambda})
\nonumber \\
& = & \frac{-i}{2\pi^2} \left [ \cos(\pi\sqrt{\lambda}(1-i)) - 
\cos(\pi\sqrt{\lambda}(1+i)) \right ]
\end{eqnarray}
according to the Weierstrass product representation.
In order to find the solution at ${\cal E}=0$, we write $\lambda = i \mu$ and get
\begin{equation}
\frac{Q_N(\lambda)}{F} = \frac{-i}{\pi^2}
 \left [ \cos(\pi\sqrt{2\mu}) - \cosh(\pi\sqrt{2\mu}) \right ]
\end{equation}
and finally, 
with  $\frac{d}{d \lambda} Q_N = \frac{1}{i} \frac{dQ_N}{d\mu}$,
 \begin{equation}
\alpha = \frac{\sqrt{2\mu}}{\pi} \frac{\cosh(\pi\sqrt{2\mu}) - 
\cos(\pi\sqrt{2\mu})}
{\sinh(\pi\sqrt{2\mu}) + \sin(\pi\sqrt{2\mu})} \; .
\end{equation}
The right-hand side of this equation is a monotonically increasing function:
 $\alpha \to \infty$ with $\mu \to \infty$.
 The dilution of the spectrum at large $|{\cal E}|$ prevents therefore
  a phase transition. \\

Fig.~6.a shows $\Gamma_k / 2$ as a function of $\alpha$ 
for $N = 50$ states.
A short-lived state is formed, but
in contrast to the ideal picket fence (Fig.~2.b),   
  {\it no} critical   value of $\alpha$ can be defined.
The width of the state in the centre of the spectrum (at ${\cal E}=0$)
increases
 smoothly as a function of $\alpha$ trapping step by step 
its  neighbours.  
The formation of the short-lived state does not occur
by a collective interaction of {\it all} basis states but by individual
trapping  of neighboured levels.
In other words,  the short-lived state is {\it not}
formed by a  cooperative effect acting over the
whole energy scale of the spectrum but is restricted to the energy range
 overlapped by it.
There is no phase transition.
\\

The number $N_{0}^p$ of principal components in the wavefunction of
the state $k=0$ is  shown as a function of $\alpha$ in
Fig.~7.a for $N= 50$ and 150. It supports the conclusion drawn.
 Also this value is a smooth
function. There is no hint to a phase transition. The broad state carries only
components of those basis states which it overlaps. In contrast to the cases
where a phase transition occurs, the slope of $N_0^p$ {\it decreases} with
increasing $N$.
The curve $N_0^p$ in Fig.~7.a remains smooth unlike
the curves for $D=0.5$ and $M=101, 301$ states
in Fig.~5.  The collective state 
is created by the basis states of a 
local energy region overlapped by it.
\\ 

\subsection{System with unequally distributed  levels and unequal coupling 
strength}

We investigate now the question whether  a  system with
diluted level density  at large $|{\cal E}|$ 
 shows a phase transition  if the states 
 at the border are coupled stronger to the decay channel than those 
in the centre of the spectrum.\\

To this purpose we rewrite Eq.~(\ref{eq:CPga}),
\begin{equation}
\alpha = \frac{-i}{ -\frac{|v_{0}|^2}{\lambda} + \sum_{k=1}^N
\left [ \frac{|v_{k}|^2}{E_k-\lambda} + \frac{|v_{-k}|^2}
{E_{-k}-\lambda} \right ] } \; .
\label{eq:a1}
\end{equation}
As in the foregoing examples, the  spectrum is supposed to be 
symmetrical (or nearly symmetrical) in relation to
 $E_{k=0}=0$:
$-E_k\approx E_{-k}$ und $v_k\approx v_{-k}$.
Looking at 
the solution at $E=0$, we write
$\lambda=-i\mu$. Then we get  
\begin{equation}
\alpha \approx \frac{-1}{ \frac{|v_{0}|^2}{\mu}+ \sum_{k=1}^N
\frac{2\mu |v_k|^2}{E_k^2 + \mu^2}  }
\label{eq:a2}
\end{equation}
from Eq.~(\ref{eq:a1}).
Since we are interested in the question whether the system shows a phase
transition at a finite value of $\alpha$, we have to consider
 (\ref{eq:a2}) in the limiting case $\mu \to \infty$ and
$N\rightarrow\infty$:
\begin{equation}
\frac{1}{\alpha_{{\rm crit}}} = \lim_{\mu\to\infty}\left [ \lim_{N\to\infty}
2 \mu \sum_{k=1}^N \frac{|v_k|^2}{E_k^2 + \mu^2} \right ] \; .
\label{eq:a3}
\end{equation}
We approximate Eq.~(\ref{eq:a3}) by an integral,
replace the discrete index $k$ by the continuous variable $x$ 
and assume
$E_k^2\approx x^t$ und $v_k^2\approx x^r$. Then
\begin{eqnarray}
\frac{1}{\alpha_{{\rm crit}}} = \lim_{\mu\to\infty} 2\mu \int_0^{\infty}
\frac{x^r}{x^t + \mu^2} dx
= \lim_{\mu\to\infty} 2\mu^{2\frac{r+1}{t}-1} \int_0^{\infty} 
\frac{s^r}{s^t +1} ds \; .
\label{eq:inta}
\end{eqnarray}
The integral converges when 
$t=r+1+\varepsilon \;\; \forall \varepsilon>0$.
\\

Let us consider the following cases.
\begin{enumerate}
\item
 $2(r+1)>t$. In this case,   $\alpha_{{\rm crit}} \to 0$ 
with $\mu \to \infty$.\\
The system is in an overcritical situation for all $\alpha > 0$. 
A phase transition does therefore {\it not} take place.

\item
$2(r+1)=t$. In this case, $\alpha = \alpha_{{\rm crit}}> 0$ remains finite  
in the limiting case $\mu \to \infty$. \\
For $\alpha<\alpha_{{\rm crit}}$, the widths of all states increase. At 
$\alpha=\alpha_{{\rm crit}}$, a phase transition {\it takes place}: the short-lived
state appears suddenly and a clear separation of time scales, with respect
to the lifetimes of the resonance states arises, even if the spectrum is
infinitely extended.

\item
 $2(r+1)< t$. In this case, $\alpha_{{\rm crit}} \to \infty$ with $\mu \to \infty$.\\
For all finite values of $\alpha$, there
exist  states  which are not overlapped 
by the collective resonance,
 whose widths  increase with increasing $\alpha$.   
Therefore, the state at ${\cal E}=0$ traps new states endlessly.
As a consequence, the formation of the short-lived state at ${\cal E}=0$ takes place 
smoothly. A phase transition does {\it not} take place.

\end{enumerate}

This analytical study 
shows the following result.
 To fulfill the conditions for a phase transition, the energy dependence of
the unperturbed spectra of ${\cal H}^0$ must be compensated by an energy 
dependent coupling of the individual states to the decay channel.
 A phase transition exists in the cases considered,
if the energy dependence of the  distribution
of the levels $E_k$  
is opposite to that of the coupling matrix elements $v_k$.
According to 
 Eq.~(\ref{eq:inta}), the critical value of $\alpha$ is
\begin{eqnarray}
\alpha_{{\rm crit}} = \frac{r+1}{\pi} = \frac{t}{2 \pi}\; .
\label{eq:critval}
\end{eqnarray}
 Otherwise, the system is either in an overcritical regime
(corresponding to $\alpha_{{\rm crit}} \to 0$) or in an undercritical one 
(corresponding to $\alpha_{{\rm crit}} \to \infty)$.
\\

Further the width of the broad pole at $\alpha\approx\alpha_{{\rm crit}}$ in the 
compensated case $(r\ne 0)$ can be estimated with respect to the length of
the spectrum.
The characteristic polynomial (Eq.~(\ref{eq:CPga})) at the energy of the collective state
$({\cal E}=0)$ reads:
\begin{eqnarray}
\frac{-i}{\alpha}
= \sum_{k=-N}^N \frac{|v_k|^2}{E_k + \frac{i}{2} \Gamma}
= \sum_{k=-N}^N \frac{|k|^r \; (k \; |k|^r -\frac{i}{2} \Gamma)}{(k^{r+1})^2 + \Gamma^2/4}
\end{eqnarray}
The principal value of that sum gives zero. Therefore we can write:
\begin{eqnarray}
\frac{1}{\alpha} & = & \frac{\Gamma}{2} \sum_{k=-N}^N \frac{|k|^r}{(k^{r+1})^2
+
 \Gamma^2/4}
\nonumber \\
& \approx & 
\frac{\Gamma}{2} \int_{-N}^N \frac{|x|^r}{(x^{r+1})^2 + \Gamma^2/4} dx 
 = \frac{\Gamma}{r+1} \int_{0}^{N^{r+1}} \frac{ds}{s^2 + \Gamma^2/4} 
\nonumber \\
& & = \frac{2}{r+1} \int_0^{\frac{2 N^{r+1}}{\Gamma}}
 \frac{d(\frac{2S}{\Gamma})}{(\frac{2S}{\Gamma})^2 + 1}
= \frac{2}{r+1} \arctan \bigg( \frac{2 N^{r+1}}{\Gamma}\bigg)
\end{eqnarray}
Using the expression (\ref{eq:critval}) for the critical point of the infinite system one gets:
\begin{eqnarray}
\frac{\pi}{2} \approx \arctan \bigg(\frac{2 N^{r+1}}{\Gamma}\bigg)
\label{eq:broadM}
\end{eqnarray}
which holds only if $\Gamma\ll 2\cdot N^{r+1}$. 
In other words, also in the compensated case,
the width of the fast decaying collective resonance state is much smaller than the extension
of the spectrum at a coupling strength close to the critical point. \\

The compensated case is illustrated in Figs.~6.b and 7.b, the overcompensated
one in 6.c and 7.c.
In the compensated case (with $r=1$), both the  distance between neighbouring levels
 and the coupling strength $|v_k|^2$
 increase linearly with $|E|$,
 ($E_k = {\rm sign}(k)\; k^2$ and  $|v_k|^2 = |k|+1 \;\; \forall k$). 
In the overcompensated case, the energy dependence of the coupling is chosen stronger
than the dilution of the spectrum ($E_k = {\rm sign}(k)\; k^2$ and $v_k =
|k|+1 \;\; \forall k$).
Here the coupling increases quadratically whereas the level density decreases linear with $E$.
\\

 The compensation of the energy dependence of the level density by 
a corresponding one in the coupling strength
restores the phase transition
(Figs.~6.b and 7.b as compared with   6.a and 7.a).
A collective mode is created 
 by  participation of (almost) all basis states. 
It occurs suddenly at a  critical value of $\alpha$.
\\

In Fig.~7.b  $N^p_0$ for $M=101, 301$ and $1001$ states is drawn.
 With increasing number $M$ of states
the curve rises  more and more sharply. The critical value is
 $\alpha_{{\rm crit}}=  2/\pi$
 (indicated by a vertical solid line) in accordance with Eq.~(\ref{eq:critval}).
\\

The maximum value $N_0^p = 1$ is not reached in Fig.~7.b. $N_0^p$ is  
approximately 0.9 and  even decreases with further increasing $\alpha$.
Drawing the contributions $b_{0i}$ in the wavefunction $\Phi_0$,
  Eq.~($\ref{eq:coeffb}$),
one sees the following feature.
The contributions
 of the states $j$  coupled more weakly to the channel 
decrease for $\alpha > \alpha_{{\rm crit}}$ in contrast to   those of the states
 $i$ coupled more strongly:  
$v_j < v_i \; \longrightarrow \; |b_{0j}| < |b_{0i}| $. Since
the differences between the $v_i$ are quite large in the cases
considered,  $N_0^p < 1$. This holds for both cases,  
decreasing and increasing energy dependence of the level density. 
In both cases, the coefficients $|b_{0j}|$ are spread at
large $\alpha$, e.g.~at
$\alpha \approx 4 \alpha_{{\rm crit}}$. This is in contrast to 
 the case of the ideal picket fence with equal coupling 
strengths, in which
 all coefficients approach the value $1 / M $
for $\alpha \ge 2 \alpha_{{\rm crit}}$.
\\

In the case of an overcompensation (Fig.~6.c and 7.c) the critical point
 is shifted to very small values
in accordance with   $\alpha_{{\rm crit}} \to 0$, Eq.~(\ref{eq:inta}). 
The number of principal components of the broad mode jumps up to $75\%$.
 Then it decreases and saturates at
around $57\%$.
\\

In an additional calculation, we bounded the spectrum from below: we
investigated the case with $E_k =  k^2, \; |v_k|^2 = k + 1 \forall k, 
\; k \ge 0 $. Also in this system, a
phase transition takes place  at $\alpha_{{\rm crit}} = 2 / \pi$ as in
the case shown in Figs.~6.b and 7.b. In all cases,  the broad
mode appears  in the energetical centre of the spectrum. \\

\subsection{System with unfolded Gaussian distributed levels}
 
It is interesting to learn whether 
the conditions for a phase transition must be fulfilled strictly or
 only on the average.
In order to answer this question in the affirmative,
we perform the following numerical analysis.
We choose an unfolded Gaussian-Orthogonal Ensemble
(GOE) for the distribution of the eigenvalues of ${\cal H}^0$ 
and a Gaussian distributed coupling
vector $V$ with mean value $\langle v \rangle =1$ and variance 
$\Delta v = 0.01$. 
\\

The decay widths as a function of the coupling parameter $\alpha$ 
are drawn in Fig.~6.d. for $N=50$.
The broad mode separates from the other ones at approximately 
$\alpha_{{\rm crit}}=1/\pi$ with a slope of $2N+1$. The features of the
 phase transition are not
as clearly pronounced in this figure as in the case of the ideal picket 
fence. Nevertheless, the differences to
Fig. ~6.a., where no phase transition occurs, are obvious. 
Even for the comparably small number of states ($M=101$),
 the fluctuations in the
distribution of the levels and the coupling vector do not destroy 
 the nature of the reorganisation process.  \\

As shown in the foregoing sections,
 the number of principal components  $N^p_k$ is a  sensitive
quantity to measure the global collectivity of the separation process.
In Fig. ~7.d.  $N^p_0$ of the collective mode at $E=0$ is 
drawn as a function of $\alpha$ for $N=50,150,250,500$. 
For increasing $M=2N+1$ the curves rise up more suddenly and the
slope  near $\alpha=1/\pi$  gets steeper. All the curves approach the maximum
value of $N_0^p$ very fast for values $\alpha>\alpha_{{\rm crit}}$. \\

The features of the second-order phase transition are better expressed
if  more resonance states
are considered. For large $N$ the irregularities in the distribution 
of the $E_k$ and $v_k$
are almost unimportant. This proves, that the conditions derived 
in the former sections 
have to be fullfilled only on the average.
Also in the ergodic case of a GOE distributed spectrum, the reorganisation 
of the spectrum can be understood as a
second-order phase transition. The fluctuations within the spectrum will be
washed out if the conditions for a phase transition derived from 
 Eq.~(\ref{eq:inta}) are fulfilled on the average.
\\

\subsection{System with complex coupling parameter $\bf \alpha$}

Up to now, we considered the system to be described by the Hamiltonian
${\cal H}$,   Eq.~($\ref{eq:ham}$),
where the coupling between system and continuum is
supposed to be real and ${\cal H}$ is non-hermitian. There may be an
additional part $\beta \; \tilde V \tilde V^+$ in the coupling term
   by which a collective state of another (internal) type  is
created. This collective state is shifted by  an energy  $\Delta {\cal E}$   from
the group of the remaining $N-1$ states \cite{sorosamu}. The structure of both
parts $V V^+$ and $\tilde V \tilde V^+$ is the same. The difference is the
non-hermiticity of the external coupling term
in the first case and the hermiticity in the second case.\\

We are interested in the question whether
 the additional term
has an influence 
 on the phase transition. Investigating this question,
we restrict ourselves   to the case $V V^+ = \tilde V \tilde V^+$, i.e.
an angle zero between the vectors $V$ and $\tilde V$.
Further, the characteristic polynomial
 (\ref{eq:CP}) does not contain, in the
one-channel case, the phases of the coupling matrix elements, but only
$|v_k|^2$.   It is justified, therefore, to replace $\alpha$ by $\alpha + i
\beta$ in the equations considered in the previous sections in order to
get conclusions on the influence of the term with $\beta \ne 0$   on the
phase transition.\\

Considering the picket-fence model with equal coupling strength
(which we studied in section 3 for $\beta = 0$)
Eq.~(\ref{eq:CPtrig}) must be replaced by
\begin{equation}
P(\lambda) = \sin(\pi\lambda) + i \pi (\alpha + i\beta) \cos(\pi\lambda)
\label{eq:pfak} 
\end{equation}
Using the representation  $\lambda = {\cal E}-\frac{i}{2}\Gamma$, one gets
\begin{eqnarray}
\left (
\begin{array}{cc}
(e^{\pi\Gamma}-1)-\pi\alpha(e^{\pi\Gamma}+1) & -\pi\beta (1-e^{\pi\Gamma}) \\
-\pi\beta (1+e^{\pi\Gamma})  & (e^{\pi\Gamma}+1) -\pi\alpha (e^{\pi\Gamma}-1)
\end{array}
\right ) {\cos(\pi E)\choose \sin(\pi E)} = 0 \; .
\label{eq:CPmat}
\end{eqnarray}
This equation has a solution,
when the determinant of the
matrix vanishes. This condition gives
\begin{equation}
\Gamma = \frac{1}{2\pi} \ln\left (\frac{(\pi\alpha +1)^2 + (\pi\beta)^2}
{(\pi\alpha -1)^2 + (\pi\beta)^2}\right ) \; .
\label{eq:gacrit}
\end{equation}
Eq.~(\ref{eq:gacrit})
 has no singularity when $\beta \ne 0$. This means, the 
 singularity in the widths of the resonances, obtained for $\beta = 0$ 
in Eqs.~(\ref{eq:CPri})
vanishes when $\beta \ne 0$.\\

This result can be understood as follows.
 In \cite{hemuro} the distribution
of exceptional points is investigated for systems 
described by
a Hamiltonian of the type $\tilde H = H^0+\tilde\alpha H^1$. 
The 
exceptional points of such a system are those points in the parameter space of
 $\tilde\alpha$
at which two (or more) eigenvalues coincide
(for a more detailed discussion see for example \cite{he}).
The coupling constant 
$\tilde\alpha$ can be
a real,  imaginary or, more generally, a complex number.
The distribution of
 the exceptional points
is determined by the matrices $H^0$ and $H^1$. It is independent of the value
 of $\tilde\alpha$ which determines, for its part,
 the positions of the (in general complex) eigenvalues of $\tilde H$.
In the one-channel case (in which $H^1$ has rank one), there exist $M-1$
exceptional points corresponding to the crossing of the collective state
with each of the other $M-1$ states.
\\

For systems which show a phase transition, {\it all}  $M-1$ exceptional points 
converge to the finite {\it purely real} value of $\tilde\alpha
= \tilde\alpha_{{\rm crit}}$ in the limit $M\to\infty$. 
For finite systems, almost all  exceptional points are near to this
accumulation point.
This result holds not only for the ideal picket fence, but for
{\it all} systems, which suffer a phase transition \cite{hemuro}. 
\\

For the systems investigated in the present paper it  follows. 
The accumulation point being  determined by the 
matrices $H^0$ and $VV^T$, is {\it independent} of the coupling parameter $\alpha$. 
When the system with
 a purely imaginary coupling  (i.e. $\beta=0$) shows a phase transition,
 {\it all} $(M-1)$ 
exceptional points are met if $\alpha$ 
approaches the critical value.
The collective mode repels with all the other ones simultaniously,
 i.e.~all states run through their exceptional point
at $\alpha=\alpha_{{\rm crit}}$.
 This means all $M-1$ exceptional points
are  accumulated at $\alpha_{{\rm crit}}$.
In that case, $\langle\Phi_i|\Phi_i\rangle $ diverges for
all $i$ simultaneously in the limes $\alpha \to \alpha_{{\rm crit}}$.
In fact, the dimension of the eigenspace collapses from $N$ to 1
$\; (|\Phi_i\rangle=|\Phi_j\rangle \; \forall i,j)$ if $\alpha$ hits the accumulation point.
Therefore, also $B$ diverges at $\alpha_{{\rm crit}}$.
If  the coupling parameter is complex ($\beta\ne 0$), however,
 the system passes the accumulation point in a certain distance
in the complex $(\alpha,\beta)$-plane. 
As a result, the singularity at $\alpha_{{\rm crit}}$ will be avoided.
For $|\alpha| \approx |\alpha_{{\rm crit}}|$,
the quantity $B$ does not diverge if $M \to \infty$
but reaches a certain maximum value.
Refering to this result, we claim that, 
according to a rigorous mathematical definition,
 the phase transition  will be destroyed by  any given 
 nonvanishing real part $\beta$
in the coupling parameter (a detailed discussion of this
aspect is given in \cite{hemuro}).\\

Let us illustrate this  result by means of a numerical study.
To that purpose, we replace $\alpha$ by $\alpha \cdot e^{i \varphi}$. We
choose $M = 2N+1 = 101$, as usually, and perform the calculations 
for  $\varphi = 1^0, 10^0, 45^0, 80^0, 89^0$ 
by varying $\alpha$.
The eigenvalues of ${\cal H}^0$ (for $\beta = 0$)
and the coupling matrix elements 
are chosen to be  $E_k=k$ 
and $v_k=1$, respectively. In Fig.~8.a.~we have drawn the number of principal components
of the collective resonance state as a function of $\alpha$.\\

The numerical results show a clear difference between the
cases with small and large angle $\varphi$.
The larger  $\varphi$, the less 
is the number of basis states contributing to the collective state at a given
$\alpha > \alpha_{{\rm crit}}$.
Further, the curves rise up more smoothly
when $\varphi$ is larger. For large angles the maximum value
$N_0^p = 1$ is not reached at the maximum value $\alpha=2$ shown in the figure.
Thus, in the case of the finite spectrum studied numerically,
 the reorganisation process is getting smoother
the larger $\varphi$. 
 In other words, the reorganisation
process is washed out, if the system cannot hit
the accumulation point  of the
exceptional points, but has to pass it in a certain distance
in the complex parameter space.
\\

This conclusion is underlined by the results given in Fig~8.b.
Here we have fixed the angle $\varphi$ to  $45^0$ and varied the number of states
included in the spectrum ($M=101$ and $1001$ states). 
As one can see, the characteristic features of the curves are not 
changed by changing the number
of basis states. The transition remains smooth
also for  $N = 1001$ states
over the whole range of $\alpha$. 
So, these systems can not be characterized by a
critical point.  We see  a critical region of $\alpha$, which will be the larger, 
the larger the nearest distance between the accumulation point and the 
 eigenvalues in the complex plane is.
\\

\section{Discussion of the results}

The analytical and numerical investigations
represented in the foregoing sections point to
similarities and differences in
the behaviour of the different systems under the influence of varying
strength of the coupling  to the continuum (decay channel).
In any case, a restructuring    in the system takes place
(or starts to take place) when the
coupling parameter $\alpha$ is large enough.
A collective state 
which is aligned with the decay channel
is formed in the centre of the spectrum. 
Its wavefunction is coherently mixed in the set of basis
wavefunctions  of the corresponding closed system.
The trapped  states  have incoherently mixed
wavefunctions.  Beyond a certain value of the control parameter
$\alpha$, two different time scales exist ({\it bifurcation of the
widths}). \\ 

In some
cases, the restructuring in the  system can be identified as  a second-order phase
transition. The separation of different time scales occurs suddenly
at a critical value $\alpha_{{\rm crit}}$ and
is a collective effect of the {\it whole} spectrum. 
The order parameter $\Gamma_0/M$ increases linearily as a function of
the control parameter $\alpha$ with a universal slope 
one, as soon as the control parameter
is larger than its critical value.
In other cases,  the
separation of time scales occurs  successively by individual trapping of
neighbouring 
resonance states. In that case, the collectivity is restricted by
the extension of the energy region, overlapped by the fast decaying 
resonance state.
The process of reorganisation continues up to $\alpha \to \infty$.  \\

The differences in the behaviour of systems which show a phase transition
to those which do not, can be seen nicely in the example of the ideal picket fence. 
In the case with equally distributed levels coupled with the same
strength to one common channel, all states are equivalent. 
Consequently, the direction of the energy shift
accompanying the local resonance trapping is undefined and the local
resonance trapping is hindered. The redistribution of the system under the
influence of the decay channel can take place only collectively.
The quantity $\Gamma_0 / M$ rises linearly in 
$\alpha > \alpha_{{\rm crit}}$ with slope 1.  
\\

More realistic systems  show a phase
transition when the energy dependence of the level density  is compensated 
by an energy dependence of the coupling strength. For example,
a dilution of the level density
can be compensated by a corresponding enhancement of the coupling strength. 
The critical value $\alpha_{{\rm crit}}$ is well determined.
Further when the system is bounded from below, 
a phase transition occurs under the same conditions as for non-bounded 
systems. It occurs at the same critical value $(r+1)/\pi$.\\

Moreover we could show, that the conditions for a phase transition do
not have to be fulfilled strictly, but only on the average. Small irregularities
in the energy dependence of the levels or in the distribution of the coupling vectors
are washed out, if the number of states in the spectrum is sufficiently high.
Also the generic case of GOE-distributed states shows the features of a phase transition
even for a comparably small number of states. The fast decaying state is
created by {\it all} states of the spectrum, independently of whether 
they are overlapped
by it or not. 
\\

Another characteristic feature of the phase transition is the 
mixing  of the wavefunctions. In the case of the ideal 
picket fence, it changes suddenly
at $\alpha_{{\rm crit}}$  from its minimum value
 $N_0^p = 1 / M$ to the maximum value $N_0^p = 1$ 
for the state $i=0$. 
The width of the 
collective resonance state at $\alpha=\alpha_{{\rm crit}}$
is of the order of $\ln(M)$ whereas the extension of the spectrum is
 equal to $M$.
Also in the more general case, where the energy dependence of the level density
is compensated by the coupling strength of the resonance states, we could prove, that the
width of the collective state is much smaller than the extension of the spectrum
for couplings close to the critical point.
Nevertheless, the collective state carries contributions of (almost)
 {\it all} 
basis states even if they are,
in the case of a finite spectrum, close to the borders.
When  $\alpha \ne \alpha_{{\rm crit}}$,
the value of
$N_0^p$  is independent of $\alpha$.   
This holds also if the system is disturbed by random pertubations, where the compensation conditions
are fulfiled only on the average.
\\

The situation is different if the energy dependence of the level density 
is not compensated by the energy dependence of the coupling vector. 
In such a case, the local resonance
trapping between neighbouring states is not hindered but occurs
successively starting in the region with the largest level density
or coupling strength.
If the level density is larger in the centre of the spectrum than at
other energies and the coupling strength to the channel is the same
for all levels,
the collective mode is created only locally for any finite value of $\alpha$.
This local collective state traps successively more and more resonance
states in direction to the border of the spectrum.
This process of local resonance trapping with increasing width
of one state in the centre continues endlessly
up to $\alpha \to \infty$  (in the limit $M \to \infty$).
Although the short-lived state
has collective properties it is not created by all basis states of the spectrum
but only by those which are overlapped by it.
There is no phase transition at all. \\

The  resonance structure of the system
is, in the cases considered,  symmetrical in relation to the   
critical value
$\alpha_{{\rm crit}}$ of the control parameter 
although the number of long-lived
resonance states for $\alpha < \alpha_{{\rm crit}}$ and for $\alpha > \alpha_{{\rm crit}}$
differs by 1.  As an example,  the picket-fence  distribution
with   level distance $1$ and equal coupling strength to the continuum
remains a picket-fence distribution also at
$\alpha > \alpha_{{\rm crit}}$ but is shifted in energy
by  1/2 of the level distance.  \\

For finite $M$, collective states may be caused also by an additional 
real part to the
Hamiltonian, e.g. ${\cal H}' = {\cal H}^0 + \beta V V^+$,
leaving ${\cal H}'$ hermitian. 
It is called internal collectivity in contrast to the external collectivity discussed above.
In such a case, the process of
reorganization in the system occurs smoothly. A phase transition does not
take place.  The eigenfunctions of 
${\cal H}'$ are orthogonal in the usual manner:
$\langle \Phi_i'| \Phi_i'
 \rangle = 1$ for all $i$ and $\alpha$.\\

Also the non-hermiticity of the Hamiltonian ${\cal H}$
is, however, not sufficient for the appearance of a phase transition,
as the results presented in the foregoing sections show.
 In any case, the value  $B$ characterizing the 
 bi-orthogonality of the set of eigenfunctions
of ${\cal H}$ plays a decisive role. Only when
it becomes essentially, i.e. when
$B \gg 1$ (see Eq.~(\ref{eq:nonher4}))
at a certain well-defined  value of $\alpha$, a
phase transition takes place.
    When, however,
the reordering of the system takes place
successively in a limited region of the spectrum 
with $n \ll M$ states, then
$B^{(n)} \equiv \frac{1}{n} \sum_{i=1}^n
 \langle  \Phi_i| \Phi_i \rangle >1$ but  $B$
 is close to 1. In this case, the reorganisation in
 the system does not occur collectively but smoothly as a function of
 $\alpha$.  \\

As a result, in the case of a  phase transition
 the bi-orthogonality of {\it all} the  eigenfunctions
  of ${\cal H}$  is maximal at (almost) the same  value of
 $\alpha$ and, according to
Eq.~(\ref{eq:nonher4}), $B \gg 1$
at $\alpha_{{\rm crit}}$. 
For illustration we show in Fig.~9 the value of
$B$ as a function of $\alpha$  for four different cases.
The theoretical value of $\alpha_{{\rm crit}}$ is  marked by a 
vertical solid line.
Only in the case without phase transition, this sum is always very close to 1  
while it has a clearly expressed maximum at the critical point $\alpha_{{\rm crit}}$
whenever a phase transition occurs.
Further, the eigenfunctions of ${\cal H}$ are orthogonal in the usual manner
for $\alpha \ll \alpha_{{\rm crit}}$ as well as for $\alpha \gg \alpha_{{\rm crit}}$.
Here, $B \approx 1$.\\

This result can nicely be illustrated
by means of the exceptional points defined as the crossing points
of resonance states. They are determined by the structure of the
 different parts
 of the effective Hamiltonian (see sect. 4.4)
but are independent of the coupling parameter. In the one-channel
case considered by us, they accumulate at  one point
in the limit $M \to \infty$. 
In the case the system shows a phase transition, the eigenvalues
meet this accumulation point when considered as a function 
of the coupling parameter and $B \to \infty$ in its neighbourhood.  
\\

Therefore, we may differentiate between four situations: (i) the exceptional
points accumulate at a finite real value in the complex $(\alpha,
\beta)$-plane {\it and} the system goes through the accumulation 
point, (ii) the exceptional
points accumulate at a finite real value in the complex $(\alpha,
\beta)$-plane but the phase $\varphi$ of the coupling hinders the system
to hit the accumulation point, 
(iii) the exceptional points accumulate at $\alpha, \beta = 0$,
(iv) the exceptional points do not accumulate
at all but they are  spread  over the whole complex  $(\alpha,
\beta)$-plane with diverging absolute value of the coupling parameter.
 Examples are (i) the compensated case, (ii) the system with
 complex coupling, (iii) the overcompensated case,
(iv) the undercompensated case (for details see \cite{hemuro}).\\

The stochastic processes
described by the (complex) partial widths are  much larger
 for $\alpha \approx \alpha_{{\rm crit}}$ than at other values of $\alpha$. This is 
expressed by the relation $\Gamma_i = \gamma_{ic}  /  \langle \Phi_i | 
\Phi_i \rangle$ where $\gamma_{ic}$ is the partial width of the state $i$ 
in relation to the (only) decay channel $c$ \cite{ro91}. 
Since     $\langle \Phi_i | \Phi_i \rangle \gg 1$ near $\alpha_{{\rm crit}}$, 
it follows $\Gamma_i \ll \gamma_{ic}$ for $\alpha \approx \alpha_{{\rm crit}}$.
The structure observed in the cross section is determined by the 
$S$-matrix, Eq.~(\ref{eq:smat}). 
 It depends essentially,  according to
 Eq.~(\ref{eq:smat}), on the 
length of the spectrum (i.e on the  values 
$\gamma_{ic}$ of all the states), but not on the width $\Gamma_{i=0}$ 
of the collective state. 
For illustration, $\overline {|1 - S_{11}|^2}$ is shown in Fig.~10
for three different values of $\alpha \geq \alpha_{{\rm crit}}$.
The width of the collective state is  $\Gamma_0 / 2 = 0.84,
~1.09$ and $~19.9$, respectively, for the three values of $\alpha $ 
considered.
The width $\Gamma_0 $ has almost nothing in common with  the structure 
observed 
in the cross section as one easily sees from the figure. 
\\

The same result follows also from our analytical considerations. 
At the critical point, the sum of the widths of all states is smaller than the
total length $M=2N+1$ of the spectrum by a factor $\pi$
according to Eq.~(\ref{eq:summenf}). Furthermore, the width
of the broadest state (in the centre of the spectrum) is in the order of 
$\ln  N$ in the ideal picket-fence model. In the limit of large $N$ it 
is in general tiny compared to the length of the spectrum according to
Eq.~(\ref{eq:broadM}).
This shows, that the transition is caused by the cooperative
behaviour of all states. 
It is {\it not} caused by the   
  overlap of the complete
spectrum by one of the states.
In the cross section, we see a structure of the extension of the length of the
spectrum since {\it all} states are coupled to the decay channel.
The width of the broadest state is much smaller than this structure.
\\

The numerical results show further that the
 number $M$ of states need not 
 necessarily be infinite in  deciding 
 the question whether the transition is of second order or not. The
 second-order phase transition is well expressed already for a relatively
 small number of states ($M=2N + 1 = 101$ up to $1001 $ in our calculations)
  in all cases  in which
 the analytical study shows a phase transition  in the limit $M \to \infty$.
   \\

In  our analytical and numerical studies  the limiting case
$M \to \infty$ is achieved by an extension of the length of the spectrum. 
It is worthwhile to note that the results are the same if, instead, the length 
of the spectrum is kept fixed at some finite value  and the level density 
approaches $\infty$ with $M \to \infty$.\\

In  many-particle systems, the level density
depends on  energy. In nuclei, it increases exponentially with energy. The
coupling strength of the states to the continuum decreases, however, with
energy due to the  increasing contribution of many-particle many-hole
configurations to the wavefunctions of the states
(''compound nucleus states''). It is an interesting question whether in
 such a system the increasing level density is
  ''compensated'' in a certain energy
 range of the spectrum by the   decreasing
 mean coupling strength so that the condition for a second-order phase
 transition is fulfilled.      \\

\noindent
{\small {\bf Acknowledgment:} We gratefully acknowledge valuable discussions 
with  F.~Leyvraz.  
C.~Jung thanks CONACYT for a beca patrimonial. The work is supported by DAAD
and DFG.}

   \vspace{2cm}

\section*{Figure captions}

\noindent
{\bf Fig.~1}.\\

\noindent
Numerical illustration of
  Eq.~(\ref{eq:eSimag})
for different values of $N$ and $\Gamma$ in the positive energy range.
{\bf (a)} $\Gamma = 1, N= 500$ (full line), 1000 (dashed line), 5000 
(dash-dotted line),
 {\bf (b)} $N = 1000, \Gamma = 0.5$ (full line), 0.75 (dashed line), 1.0
(dash-dotted line). For details see text.\\

\noindent
{\bf Fig.~2.}\\

\noindent
{\bf (a)} The motion of the eigenvalues $\lambda_k$ in the complex plane 
with increasing $\alpha$ for the ideal picket fence ($N=50$).
The ordinate represents the values of $\Gamma_k / 2$ while the abcisse
those of ${\cal E}_k$. The full line shows the behaviour of 
Eq.~(\ref{eq:huell}). Only a part of the spectrum is shown.
{\bf (b)}  $\Gamma_k / 2$ as a function of $\alpha$ for $N=50$. 
{\bf (c)} $N_0^p$ as a function of $\alpha$ for $N=50$ (full line), $150$
(dotted line).
\\

\noindent
{\bf Fig.~3}\\

\noindent
{\bf (a)}
Eigenvalues $\lambda_k = {\cal E}_k - i/2 \; \Gamma_k$ in the 
complex plane for a small energy range around
the centre. 
$E_k = k \; \forall k$ and $v_k = 1 \; \forall k$ but $v_0 = 0.5$. 
 $N=50$ (rhombs), 150 (crosses). 
{\bf (b)} $\Gamma_k$ as a function of $ (\alpha)$.
$E_k = k \; \forall k$ and $v_k = 1 \; \forall k$ but $v_0 = 0.5$. 
   $N= 50, 150$. 
\\

\noindent
{\bf Fig.~4}\\

\noindent
Numerical illustration of Eq.~(\ref{eq:CPoTItri}).
$\pi\mu \coth(\pi \mu) + D$ 
(left ordinate scale)
for $D = -0.5, 0, 0.5$ (full lines)  
and $\mu / \alpha$ (right ordinate scale)
for different $\alpha$.
{\bf (a)} for $\alpha = 0.1$ (dashed line), $1/\pi$ (solid thin line). 
{\bf (b)} for $\alpha_{{\rm crit}}<\alpha<\alpha_{c2}$ (solid thin line), 
$\alpha\approx\alpha_{c2}$ (dashed line),
$\alpha > \alpha_{c2}$ (dash-dotted line).
For details see text.\\

\noindent
{\bf Fig.~5} \\

\noindent
Number $N^p_0$ of principal components as a function of $\alpha$
 for the three different picket-fence
distributions with $v_0 = 0.5, N=50$ (dashed line), $v_0 = 1, N=50$ (full line),
and $v_0 = 2, N=50$ (dash-dotted line), 150 (double-dotted-dashed line).\\

\noindent 
{\bf Fig.~6}\\

\noindent
$\Gamma_k / 2$ as a function of $\alpha$ for 
$N=50 $.
{\bf (a)} $E_k = {\rm sign} (k) \; k^2, \; v_k = 1 \; \forall k$,
{\bf (b)} $E_k = {\rm sign} (k) \; k^2, \; v_k^2 = |k| + 1 $,
{\bf (c)} $E_k = {\rm sign} (k) \; k^2, \; v_k = |k| + 1 $,
{\bf (d)} $E_k$ distributed according to an unfolded GOE, $v_k$ are Gaussian
distributed with mean value 1 and variance 0.1.
  \\

\noindent 
{\bf Fig.~7}\\

\noindent
$N^p_0$ as a function of $\alpha$ for 
{\bf (a)} $E_k = {\rm sign} (k)\; k^2,\; v_k = 1 \; 
\forall k$, $N = 50$ (full line),
150 (dashed line);
{\bf (b)} $E_k = {\rm sign} (k) \; k^2,\; v_k^2 = |k| + 1 $, $N=50$ (full line),
150 (dotted line), 500 (dashed line);
{\bf (c)} $E_k = {\rm sign} (k)\; k^2,\; v_k = |k| + 1 $, $N=50$ (full line);
{\bf (d)} $E_k$ distributed according to an unfolded GOE, $v_k$ are Gaussian
distributed with mean value 1 and variance 0.1.
$N=50 $ (full line), 150 (dotted line), 250 (dashed line), 500 (dash-dotted 
line).
\\

\noindent 
{\bf Fig.~8}\\

\noindent
$N^p_0$ as a function of $\alpha$ for a system with complex coupling.
Simultaneously, $\beta$  changes according to $\beta = \alpha \tan (\varphi)$.
{\bf (a)} 
The values $\varphi = 1^0, 10^0, 45^0, 80^0, 89^0$ are shown as full,
long-dashed,
short-dashed, dotted, dash-dotted line, respectively. $N=50$.
{\bf (b)} $\varphi = 45^0, N = 50$ (full line), $N=150$ (dotted line).
\\

\noindent
{\bf Fig.~9}\\

\noindent
The value of $B$,
 Eq.~(\ref{eq:nonher4}), as a function of $\alpha$.
The thick full line shows the case $E_k = {\rm sign} (k) \; k^2,\; v_k =1
\; \forall k, N = 50$, 
thin solid  line  the ideal picket fence for $ N = 50$,
dashed  line  the ideal picket fence for $ N = 150$,
dash-dotted  line the unfolded GOE as in Fig.~7.d for $ N = 150$.\\

\noindent
{\bf Fig.~10}\\

\noindent
$\overline {|1 - S_{11}|^2}$ for  $\alpha = 1 / \pi$ (full line), 
$1 / \pi + 0.05$ (dashed line), $1 / \pi + 0.1$ (dash-dotted line).
For comparison: $\Gamma_0 / 2 = 0.84$  for  $\alpha = 1 / \pi$,
 $\Gamma_0 / 2 = 1.09$ for $\alpha = 1 / \pi + 0.05$ and $\Gamma_0 / 2 
= 19.9$ for $\alpha = 1 / \pi + 0.1$.

\end{document}